\documentclass[fleqn,usenatbib]{mnras}

\usepackage{newtxtext,newtxmath}

\usepackage[T1]{fontenc}

\DeclareRobustCommand{\VAN}[3]{#2}
\let\VANthebibliography\thebibliography
\def\thebibliography{\DeclareRobustCommand{\VAN}[3]{##3}\VANthebibliography}

\usepackage{graphicx}	
\usepackage{amsmath}	
\usepackage[flushleft]{threeparttable}

\newcommand{\MgIIdblt}{{\rm Mg}\kern 0.1em{\sc ii}~$\lambda\lambda 2796, 2803$}
\newcommand{\MgI}{\hbox{{\rm Mg}\kern 0.1em{\sc i}}}
\newcommand{\MgII}{\hbox{{\rm Mg}\kern 0.1em{\sc ii}}}
\newcommand{\MnII}{\hbox{{\rm Mn}\kern 0.1em{\sc ii}}}
\newcommand{\FeII}{\hbox{{\rm Fe}\kern 0.1em{\sc ii}}}
\newcommand{\OVIdblt}{{\rm O}\kern 0.1em{\sc vi}~$\lambda\lambda 1031, 1037$} 
\newcommand{\OVI}{\hbox{{\rm O}\kern 0.1em{\sc vi}}}
\newcommand{\CII}{\hbox{{\rm C}\kern 0.1em{\sc ii}}}
\newcommand{\CIII}{\hbox{{\rm C}\kern 0.1em{\sc iii}}}
\newcommand{\CIV}{\hbox{{\rm C}\kern 0.1em{\sc iv}}}
\newcommand{\HI}{\hbox{{\rm H}\kern 0.1em{\sc i}}}
\newcommand{\Lya}{\hbox{{\rm Ly}\kern 0.1em$\alpha$}}
\newcommand{\Halpha}{\hbox{{\rm H}\kern 0.1em$\alpha$}}
\newcommand{\Hbeta}{\hbox{{\rm H}\kern 0.1em$\beta$}}
\newcommand{\Lyb}{\hbox{{\rm Ly}\kern 0.1em$\beta$}}
\newcommand{\OI}{\hbox{{\rm O}\kern 0.1em{\sc i}}}
\newcommand{\SiII}{\hbox{{\rm Si}\kern 0.1em{\sc ii}}}
\newcommand{\SiIII}{\hbox{{\rm Si}\kern 0.1em{\sc iii}}}
\newcommand{\SiIV}{\hbox{{\rm Si}\kern 0.1em{\sc iv}}}
\newcommand{\NI}{\hbox{{\rm N}\kern 0.1em{\sc i}}}
\newcommand{\NII}{\hbox{{\rm N}\kern 0.1em{\sc ii}}}
\newcommand{\NIII}{\hbox{{\rm N}\kern 0.1em{\sc iii}}}
\newcommand{\NV}{\hbox{{\rm N}\kern 0.1em{\sc v}}}
\newcommand{\PII}{\hbox{{\rm P}\kern 0.1em{\sc ii}}}
\newcommand{\SI}{\hbox{{\rm S}\kern 0.1em{\sc i}}}
\newcommand{\SII}{\hbox{{\rm S}\kern 0.1em{\sc ii}}}
\newcommand{\AlII}{\hbox{{\rm Al}\kern 0.1em{\sc ii}}}
\newcommand{\AlIII}{\hbox{{\rm Al}\kern 0.1em{\sc iii}}}
\newcommand{\OII}{\hbox{[{\rm O}\kern 0.1em{\sc ii}]}}
\newcommand{\kms}{\hbox{km~s$^{-1}$}}

\newcommand{\etal}{et~al.}
\newcommand{\metallicity}{$\log (Z/Z_{\odot})$}
\newcommand{\hden}{$\log (n_{\rm H}/{\rm cm}^{-3})$}
\newcommand{\colden}{$\log (N(\HI)/{\rm cm}^{-2})$}

\title[Intragroup Medium of a $z=2.431$ Compact Group]{A Complex Multiphase DLA Associated with a Compact Group at $z=2.431$ Traces Accretion, Outflows, and Tidal Streams}

\author[N.~M.~Nielsen et al.]
{Nikole M. Nielsen,$^{1,2}$\thanks{E-mail: nikolenielsen@swin.edu.au}
Glenn G. Kacprzak,$^{1,2}$
Sameer,$^{3}$
Michael T. Murphy,$^{1}$
\newauthor 
Hasti Nateghi,$^{1,2}$
Jane C. Charlton,$^{3}$
and Christopher W. Churchill$^{4}$
\\
$^{1}$Centre for Astrophysics and Supercomputing, Swinburne University of Technology, Hawthorn, Victoria 3122, Australia\\
$^{2}$ARC Centre of Excellence for All Sky Astrophysics in 3 Dimensions (ASTRO 3D), Australia\\
$^{3}$Department of Astronomy \& Astrophysics, 525 Davey Lab, The Pennsylvania State University, University Park, PA 16802, USA\\
$^{4}$Department of Astronomy, New Mexico State University, Las Cruces, NM 88003, USA
}

\date{Accepted 2022 June 28. Received 2022 June 28; in original form 2021 November 25}

\pubyear{2022}

\begin{document}
\label{firstpage}
\pagerange{\pageref{firstpage}--\pageref{lastpage}}
\maketitle

\begin{abstract}
As part of our program to identify host galaxies of known $z=2-3$ {\MgII} absorbers with the Keck Cosmic Web Imager (KCWI), we discovered a compact group giving rise to a $z=2.431$ DLA with ultra-strong {\MgII} absorption in quasar field J234628$+$124859. The group consists of four star-forming galaxies within $8-28$~kpc and $v\sim40-340$~{\kms} of each other, where tidal streams are weakly visible in deep {\it HST} imaging. The group geometric centre is $D=25$~kpc from the quasar ($D=20-40$~kpc for each galaxy). Galaxy G1 dominates the group ($1.66L_{\ast}$, ${\rm SFR}_{\rm FUV}=11.6$~M$_{\odot}$~yr$^{-1}$) while G2, G3, and G4 are less massive ($0.1-0.3L_{\ast}$, ${\rm SFR}_{\rm FUV}=1.4-2.0$~M$_{\odot}$~yr$^{-1}$). Using a VLT/UVES quasar spectrum covering the {\HI} Lyman series and metal lines such as {\MgII}, {\SiIII}, and {\CIV}, we characterised the kinematic structure and physical conditions along the line-of-sight with cloud-by-cloud multiphase Bayesian modelling. The absorption system has a total $\log(N({\HI})/{\rm cm}^{-2})=20.53$ and an $N({\HI})$-weighted mean metallicity of $\log(Z/Z_{\odot})=-0.68$, with a very large {\MgII} linewidth of $\Delta v\sim700$~{\kms}. The highly kinematically complex profile is well-modelled with 30 clouds across low and intermediate ionisation phases with values ${13\lesssim\log(N({\HI})/{\rm cm}^{-2})\lesssim20}$ and $-3\lesssim\log(Z/Z_{\odot})\lesssim1$. Comparing these properties to the galaxy properties, we infer a wide range of gaseous environments, including metal-rich outflows, metal-poor IGM accretion, and tidal streams from galaxy--galaxy interactions. This diversity of structures forms the intragroup medium around a complex compact group environment at the epoch of peak star formation activity. Surveys of low redshift compact groups would benefit from obtaining a more complete census of this medium for characterising evolutionary pathways.
\end{abstract}




\section{Introduction}

Galaxies evolve by accreting metal-poor gas from the intergalactic medium (IGM), processing that gas into stars in the interstellar medium (ISM), ejecting metal-rich gas into the surrounding circumgalactic medium (CGM) via outflows, and later re-accreting the increasingly metal-rich gas from the CGM to form more stars. This transfer of gas between the IGM, ISM, and CGM, known as the baryon cycle, regulates the star formation activity within a galaxy, determining its evolutionary path from star-forming to quiescent depending on the content and properties of each of these reservoirs (e.g. \citealp{oppenheimer08, lilly-bathtub}; and reviews by \citealp{tumlinson17, perouxhowk}). Low redshift surveys of the CGM using quasar absorption lines have demonstrated that this reservoir contains over half of the baryonic mass in galaxies, which makes it a significant source of star-forming material and a key regulator of galaxy evolution \citep[e.g.][]{peeples14, werk14}. The CGM is arguably most influential at $z=2-3$ (``Cosmic Noon'') when the global star formation rate density peaks and galaxies are most actively building up their mass \citep{madau14}. The rates of accretion onto and outflows from galaxies are both predicted to be at their greatest during this epoch \citep{steidel10, rupke-review, vandevoort11}. 

The vast majority of quasar absorption line work connecting the CGM to the host galaxies has focused on ``typical'' CGM absorbers (e.g. (partial-)Lyman Limit Systems, pLLS/LLS, $16.2<\log (N({\HI})/{\rm cm}^{-2})<19.0$) at low redshift. However, the physical mechanisms and environments giving rise to some of the more rare systems such as damped {\Lya} absorbers (DLAs; $\log (N({\HI})/{\rm cm}^{-2})>20.3$) and ultra-strong {\MgII} systems ($W_r(2796)\geq3$~{\AA}) are still not well-characterised, especially at Cosmic Noon. These systems likely trace the interface between the CGM and ISM since absorption strength generally anti-correlates with impact parameter \citep[e.g.][]{chen01a, magiicat2, magiicat1, kacprzak13} and may present more extreme tests of gas flows and feedback processes because of their very large velocity widths \citep[e.g.][]{bondwinds} that are greater than are found with more typical systems. DLAs are expected to trace large amounts of neutral {\HI} in and near galaxies \citep[for a review, see][]{wolfe05} but searches for high redshift DLA host galaxies often result in non-detections \citep[for a summary, see][]{krogager17}. The galaxy counterparts for the highest metallicity DLAs are often found, but this is not the case for low metallicity DLAs with ${\rm [Zn/H]}<-1.0$. This suggests that DLAs follow a DLA metallicity--galaxy mass/luminosity relationship, where more metal-poor DLAs are hosted by less massive/luminous galaxies that are below typical survey sensitivities. In the local Universe, \citet{deblok18} created deep {\HI} emission maps around the M81 galaxy triplet (including starburst M82), where they found large {\HI} linewidths. They suggested that the superposition of outflow and tidal stream components in the system could explain the linewidths of DLAs at higher redshifts, so high redshift DLA host searches should also explore the large-scale environment \citep[e.g.][]{mackenzie19}.

Ultra-strong {\MgII} absorbers, which have $W_r(2796)>3$~{\AA} and extremely large velocity spreads of $\Delta v=320-1000$~{\kms} that cannot be explained by large column densities, can be a rare sub-class of DLAs \citep[where $W_r(2796)$ and N({\HI}) are correlated with large scatter;][]{menard09}. Previous work examining the spectroscopically-confirmed host galaxies of four ultra-strong {\MgII} absorbers, all of which are located in group environments, suggested that the absorption arose from tidal interactions \citep{gauthier13}, star formation-driven outflows \citep{nestor11}, or a merger-induced starburst \citep{rubin10}. The environments of these systems consist of several galaxies with low star formation rates at large impact parameters ($\sim40-250$~kpc) or fewer galaxies (two within $\sim60$~kpc) that are starbursting. Another ultra-strong {\MgII} absorber was associated with a very low impact parameter ($D\sim0.9$~kpc), presumably gas rich, compact dwarf galaxy that was likely undergoing a starburst \citep{noterdaeme12}. These latter authors used an observing strategy of three long-slits centred on the quasar, where they only fully surveyed the field out to $\sim10$~kpc, so they were not able to determine the environment of this galaxy. Additional information such as absorber metallicities and galaxy morphologies, which were not available for all of the literature systems, would have provided a more complete picture for interpreting the gas origins. The previous work also considered the absorption as a single structure along the line-of-sight when interpreting the gas origins. Simulations and recent photoionisation modelling methods now suggest that quasar absorption lines likely trace multiple structures along the line-of-sight \citep[e.g.][]{churchill15, peeples19, zahedy19LRG, sameer2021, sameerleo, marra22} and it is likely that these ultra-strong systems are more extreme in the number and types of structures along the line-of-sight than the typical systems studied in the simulations and previous photoionisation modelling.

It is not surprising that many of these rare {\MgII} systems have been found in group environments since stronger equivalent widths and larger covering fractions out to larger distances have been observed in group environments than found around isolated galaxies \citep{chen10a, bordoloi11, magiicat6, fossati19, hamanowicz20, dutta21, huang21}. Further characterising the kinematics of {\MgII} systems in loose groups, \citet{magiicat6} found more optical depth at higher velocity than in isolated systems, possibly a signature of an intragroup medium (IGrM) or the interactions between galaxies. A similar enhancement for small loose groups was found in {\CIV} \citep{burchett16} but {\OVI} instead becomes weaker with lower covering fractions (\citealp{pointon17, ng19, mccabe21}; although \citealp{cgm2ovi} found larger covering fractions for galaxies with neighbours and column densities $\log (N({\OVI})/{\rm cm}^{-2})<14.1$). To better understand the origin of these enhancements and complex kinematics, \citet{hamanowicz20} suggested kinematically connecting the absorption components along the line-of-sight to each galaxy in the group. Further connecting the metallicity of each component and ionisation phase to each galaxy has also resulted in new interpretations for complex absorbers that were previously thought to be associated with a single isolated galaxy \citep[e.g.][]{muzahid15, nateghi21}.

Compact groups present an interesting environment to study the IGrM since the galaxies are separated by distances comparable to the sizes of the galaxies themselves, which means that strong galaxy--galaxy interactions are taking place (interactions are not necessarily present in loose groups). They have high galaxy densities with low velocity dispersions, where the most well-known in the local Universe, Hickson Compact Groups, have median 3D separations of 70~kpc and 3D velocity dispersions of $\sigma\sim330$~{\kms} (\citealp{hickson82, hickson92}; also see \citealp{mcconnachie09} for a compilation with SDSS). These systems tend to have fewer star-forming spiral galaxies than found in the field, implying an enhanced evolution of the galaxies from star-forming to quiescent \citep{hickson82}. Optical imaging and maps of the {\HI} emission in compact groups have found extensive tidal streams from the inevitable interactions between such close galaxies \citep[e.g.][]{mdo94, verdes-montenegro01}. Perhaps unsurprisingly, these maps have also shown that the galaxies themselves have varying degrees of {\HI}-deficiencies, where the gas may have been stripped out of the galaxies, depleting their star formation reservoirs and resulting in the lower fraction of late-type spirals compared to the field \citep[e.g.][]{verdes-montenegro01}. 

Given these characteristics, \citet{verdes-montenegro01} suggested an evolutionary pathway for compact groups in which the galaxies become more quiescent as their {\HI} reservoirs are stripped from repeated interactions. The stripped gas then resides in the IGrM, where \citet{borthakur10, borthakur15} found an excess of diffuse {\HI} in increasing amounts as the galaxies become more {\HI}-deficient, which aligns with the \citeauthor{verdes-montenegro01} evolution. \citet{bitsakis16} additionally suggested that interactions between the galaxy ISM and the surrounding IGrM further suppress star formation in the galaxies, so having a census of this intragroup material is needed to determine the degree to which this mechanism plays a role in the evolution of compact groups. Studying a single compact group, \citet{konstantopoulos10} suggested that the lack of detection of a diffuse {\HI} and X-ray IGrM surrounding their galaxies implied that interactions were not taking place to redistribute ISM material into the IGrM. Instead, they suggested that the galaxies will deplete their star-forming reservoirs more quickly from star-formation than interactions will strip the gas, eventually leading to a dry merger. While the local compact groups that were used to infer these evolutionary pathways have extensive imaging, {\HI} maps, and sometimes have X-ray imaging to observe the hottest IGrM gas, they neglect the significant reservoir of $T\sim10^{3.5-5}$~K gas that is typically observed in the CGM with quasar absorption lines to much lower column densities than {\HI} imaging can reach.

Simulations have suggested that 95 percent of compact groups at $z=2$ will fully merge into a single galaxy by $z=0$ and up to $\sim15$ percent of galaxies were located in these environments in the past \citep[e.g.][]{wiens19}. Studying compact groups at earlier epochs can test these predictions. Additionally, the current compact group literature is limited to very low redshifts, which means that the accretion rate of gas onto galaxies to sustain their star formation is significantly decreased, so their evolution might simply be a consequence of their accretion being cut off \citep[e.g. cold mode accretion no longer reaches galaxies with halo masses ${>10^{12}}$~M$_{\odot}$ at low $z$;][]{birnboim03, keres09a}. Thus, studying compact groups at $z=2$ when galaxies are most actively accreting gas \citep[e.g.][]{vandevoort11} can explore whether it is the accretion cut-off or the galaxy--galaxy interactions that are causing the star formation suppression observed at low redshift. While higher redshifts do not allow for {\HI} mapping like at low redshift, they do shift the {\HI} Lyman series into the optical bands in addition to a wide range of metal ions with rest-frame UV wavelengths, all of which can be covered simultaneously in a single high resolution quasar spectrum using, e.g., VLT/UVES or Keck/HIRES to characterise the metallicities and ionisation conditions as well as obtain a census of the material in the more diffuse IGrM.

Using NIFS on Gemini-North with the ALTAIR adaptive optics and laser guide star, \citet{wang2015} searched quasar field J234628+124859 (hereafter J2346+124) for {\Halpha} emission associated with a DLA at $z_{\rm DLA}=2.431$ within a radius of $\sim 12.5$~kpc from the quasar sightline. They found no host galaxies to a limit of ${\rm SFR}>4.4$~M$_{\odot}$~yr$^{-1}$ for impact parameters $4-12.5$~kpc from the quasar and ${\rm SFR}>11$~M$_{\odot}$~yr$^{-1}$ within 4~kpc. If a DLA host galaxy is present within these impact parameters, it would have a low star formation rate that is undetectable by {\Halpha} in their observations. This suggests that either the DLA host galaxy is not strongly star-forming, or the host is beyond an impact parameter of $\sim12.5$~kpc. We re-observed this DLA, which also has ultra-strong {\MgII}, with the Keck Cosmic Web Imager \citep[KCWI;][]{kcwi} as part of our program to identify host galaxies of known $z=2-3$ {\MgII} and {\CIV} absorbers \citep{nielsen20}. With our deeper and wider survey of the field, we identified four galaxies in a compact group at the redshift of the DLA beyond the search radius of the previous survey.


This paper is organised as follows. Section~\ref{sec:methods} details our observational data and briefly summarises our reduction procedures. Section~\ref{sec:results} presents the DLA host galaxies and associated absorption, along with our methods for measuring the galaxy properties, absorption properties, and the photoionisation modelling. The results of our analysis are discussed in Section~\ref{sec:discussion}, where we speculate on the physical origins of the gas along the line-of-sight and explore the implications for compact group evolution. In Section~\ref{sec:conclusions} we summarise the results and present our conclusions. Throughout the paper we report AB magnitudes, proper distances, and adopt a $\Lambda$CDM cosmology ($H_0=70$~km~s$^{-1}$~Mpc$^{-1}$, $\Omega_M=0.3$, and $\Omega_{\Lambda}=0.7$).

\section{Data and Observations}
\label{sec:methods}

The DLA absorption was previously observed in the background quasar sightline (J2346$+$124) using high resolution Keck/HIRES and VLT/UVES spectra \citep[][and references therein]{wang2015} and the quasar field was observed with deep {\it HST}/WFC3 imaging as part of the Keck Baryonic Structure Survey \citep[KBSS, e.g.][]{rudie12, rudie19}. We used these data in combination with new Keck/KCWI integral field spectroscopy to search for the host galaxies of known $z=2-3$ absorbers within $D\sim140$~kpc of the background quasar sightline. With the quasar spectra, we modelled the multiphase gas to measure detailed gas kinematics and metallicities, while the {\it HST} imaging was used to measure galaxy photometries and morphologies in three rest-frame blue bands. The KCWI data were used to measure galaxy redshifts and star formation rates with {\Lya} and the far-UV (FUV) continuum. We describe each data set below and the methods for determining galaxy and absorption properties in Section~\ref{sec:results}. Appendix~\ref{app:gradient} additionally details our method for removing a wavelength- and spatially-dependent illumination gradient present in the KCWI data.

\subsection{Quasar Spectroscopy}
\label{sec:qsospectra}

Quasar J2346+124, $z_{\rm qso}=2.573$\footnote{There are several different redshifts listed for this quasar in NED, with values of $z_{\rm qso} = 2.515$, 2.573, and 2.763. However, a {\Lya} halo around the quasar, which is not discussed here, indicates the true redshift is $z_{\rm qso}=2.573$.} has been observed with both Keck/HIRES and VLT/UVES, but we use the UVES spectrum in the UVES SQUAD \citep{uvessquad} since it has a larger wavelength coverage than the HIRES spectra. The VLT/UVES observations consist of a total exposure time of 28200~s (PIDs 072.A-0346(A), 079.B-0469(A), 67.A-0022(A), and 69.A-0204(A)). The data were reduced using the UVES pipeline \citep{dekker-uves} and the exposures were combined using {\sc uves\_popler} \citep{uvespopler, uvessquad}. The spectrum is converted to vacuum wavelengths and corrected for the heliocentric velocity. While the pipeline fits a continuum with automated methods redward of the quasar's {\Lya} emission and by hand in the {\Lya} forest, we found that we needed to do a more careful fit around several of the absorption lines of interest, where we fitted a cubic spline to the spectrum to determine a continuum that was smooth across roughly $100$~{\kms}. The final spectrum provides full wavelength coverage of the {\HI} Lyman series, as well as metal lines up to and including wavelengths covering the {\MgIIdblt} doublet and {\MgI}~$\lambda2853$ (observed wavelengths of $3120\lesssim\lambda\lesssim10000$~{\AA}) for the DLA at $z_{\rm DLA}=2.431$.

\begin{figure*}
	\includegraphics[width=\linewidth]{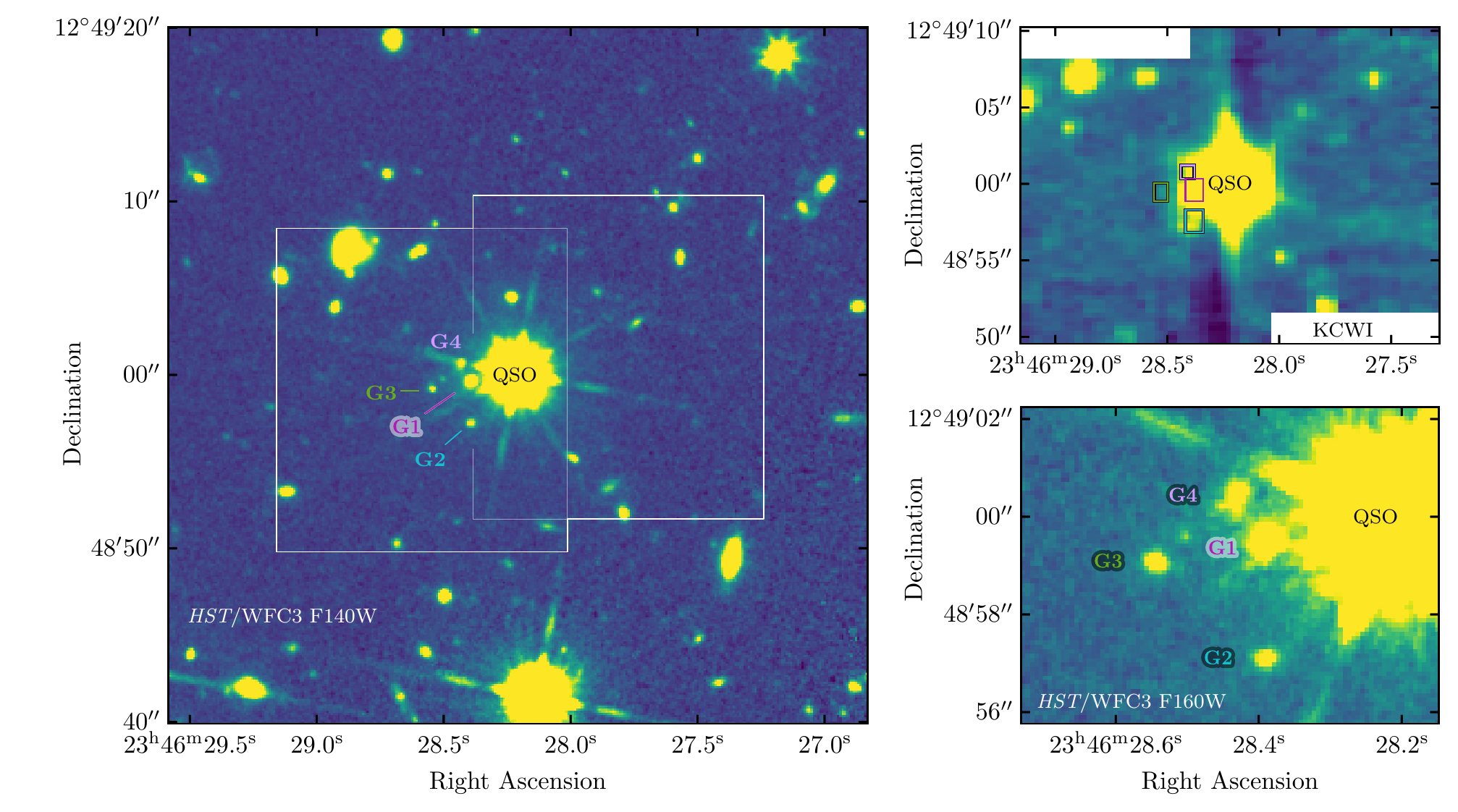}
    \caption{Overview of quasar field J2346$+$124. ({\it left}) {\it HST} F140W imaging within a $40''\times40''$ box centred on the quasar (QSO), corresponding to $D\sim160$~kpc at the redshift of the DLA, $z_{\rm DLA}=2.431$. Three images were combined with Montage for a total exposure of 7836~s. The white rectangles demonstrate the two medium slicer ($16\farcs5\times20\farcs4$) KCWI pointings used in this field. ({\it bottom right}) A zoom-in on the galaxies in the group in an {\it HST} F160W image (exposure time of 899~s), which was used for the galaxy morphology modelling. ({\it top right}) KCWI white light image centred on the quasar, where the field is surveyed out to $D=66$~kpc in all directions and with a maximum surveyed impact parameter of $D<140$~kpc at $z_{\rm DLA}$. The four galaxies in this field, G1, G2, G3, and G4 are clearly visible in the {\it HST} imaging, but are encompassed in the quasar's PSF in the KCWI white light image due to an extended {\Lya} halo around the quasar and the seeing. The galaxies are more clearly separated from the quasar in a {\Lya} line map (Fig.~\ref{fig:galspec}). North is up and east is to the left in all panels.}
    \label{fig:whitelight}
\end{figure*}

\subsection{{\it HST} Imaging}
\label{sec:HSTimaging}

J2346+124 is a well-studied field with {\it HST}, having been imaged in the F110W, F140W, and F160W filters with WFC3/IR (PIDs: 11694, 12471, 12578. and 14620). These bands cover rest-frame near-UV (NUV) and optical bands ranging from $\sim2575$~{\AA} to ${\sim4960}$~{\AA} at the redshift of the DLA. Fig.~\ref{fig:whitelight} (left) shows an F140W image (roughly rest-frame $B$-band) with a total exposure time of 7836~s centred on the quasar. For this figure, we combined three exposures of 2612~s each using {\sc Montage}.\footnote{\url{http://montage.ipac.caltech.edu/}} The bottom right panel of Fig.~\ref{fig:whitelight} also shows a small portion of the field of view with F160W, which has an exposure time of 899~s. Galaxy photometries were measured in each {\it HST} band using the Source Extractor software \citep{bertin96} with a detection criterion of $1.5\sigma$ above the background. The resulting galaxy magnitudes are quoted in the AB system.

\begin{table*}
    \centering
    \caption{DLA Host Galaxy Identifications}
    \label{tab:DLAgals}
    \begin{threeparttable}
    \begin{tabular}{lcccc}
    \hline
    Galaxy     & $z_{\rm gal}$     & R.A.         & Dec           & $\theta$ \\
               &                   &              &               & (arcsec) \\
    \hline
    G1 (West)  & $2.4306\pm0.0001$ & 23:46:28.39  &  +12:48:59.6  &  2.7 \\
    G2 (South) & $2.4322\pm0.0008$ & 23:46:28.39  &  +12:48:57.2  &  3.8 \\
    G3 (East)  & $2.428\pm0.001$   & 23:46:28.54  &  +12:48:59.2  &  4.9 \\
    G4 (North) & $2.4310\pm0.0001$ & 23:46:28.43  &  +12:49:00.6  &  3.3 \\
    \hline
    \end{tabular}
    \end{threeparttable}
\end{table*}

\subsection{KCWI Integral Field Spectroscopy}
\label{sec:KCWIdata}

We observed the J2346$+$124 field with Keck/KCWI on 2018 September 9 UT (PID: 2018B\_W232) using the medium slicer and BL grating with a wavelength centre of 4500~{\AA} and $2\times2$ binning. The medium slicer has a field of view of $16\farcs5\times20\farcs4$ and a spatial resolution of $0\farcs29\times0\farcs69$. The BL grating provides a spectral resolving power of $\sim1800$ ($\sim1$~{\AA}~pix$^{-1}$) and spans a wavelength range of $3500\lesssim\lambda\lesssim5500$~{\AA}. To survey the field, we used two pointings overlapping the area within a $>2\farcs6$ radius of the quasar, where the individual pointing footprints are highlighted as the white rectangles in Fig.~\ref{fig:whitelight} (left). Each pointing has a position angle of 0~deg and consists of three exposures of 945~s, for a total exposure of 47~min in each pointing or a total exposure of 1.6~hrs on top of the quasar and the rest of the overlap region. At the redshift of the DLA, $z_{\rm DLA}=2.431$, the field is surveyed out to a minimum impact parameter of $D\sim70$~kpc and maximum $D\sim140$~kpc, with a spatial resolution of $2.4\times5.6$~kpc.

The data were reduced using the IDL version of the KCWI Data Reduction Pipeline (DRP)\footnote{\url{https://github.com/Keck-DataReductionPipelines/KcwiDRP}} using standard settings but skipping the sky subtraction step (\texttt{kcwi\_stage5sky}). Standard star kopff27 from the KCWI DRP starlist was used to flux calibrate the data in the last step of the DRP. The data were then further reduced using in-house software as described in Appendix~\ref{app:gradient} to remove a residual illumination gradient, scattered light from the quasar, and to subtract the sky since separate sky fields were not observed. The final reduced cubes were then optimally combined using {\sc Montage} to produce square spaxels and combine all exposures for the two pointings. When reprojecting the cubes before coadding, i.e. the \texttt{mProjectCube} step, we set \texttt{drizzle=1.0} 
and scaled the flux using the \texttt{fluxScale} option by the change in the spaxel dimensions to properly redistribute the flux from rectangular to square spaxels. The variance cubes were also combined with {\sc Montage}, but the values were scaled by the change in spaxel dimensions squared and divided by the number of exposures combined, where this latter value varies across the field of view since the quasar has coverage from six exposures while most of the FOV is covered by only three exposures. This $n_{\rm exp}$ normalisation is necessary because the \texttt{mAddCube} step determines the mean flux values in each spaxel, where error propagation requires the variances of each exposure be summed and normalised by $n_{\rm exp}^2$. {\sc Montage} does not account for this extra factor of $n_{\rm exp}$ (it assumes it is working with fluxes rather than variances). All spectra are heliocentric velocity and vacuum wavelength corrected for comparison to the UVES spectra. They were also corrected for Galactic extinction using the \citet{sf11} dust reddening map and assuming a \citet{cardelli89} attenuation law. The final data cube has an average $3\sigma$ surface brightness limit (the objects are masked) of $1.4\times10^{-18}$~erg~s$^{-1}$~cm$^{-2}$~arcsec$^{-2}$, which corresponds to ${\rm SFR}_{\rm FUV}\gtrsim1.4$~M$_{\odot}$~yr$^{-1}$ assuming a $1''\times1''$ region.

The KCWI white light image is plotted in the upper right panel of Fig.~\ref{fig:whitelight}, where the wavelength axis has been collapsed. The quasar (QSO) is the brightest object at the centre and many continuum objects corresponding to galaxies in the {\it HST} image are present.

\begin{figure*}
	\includegraphics[width=\linewidth]{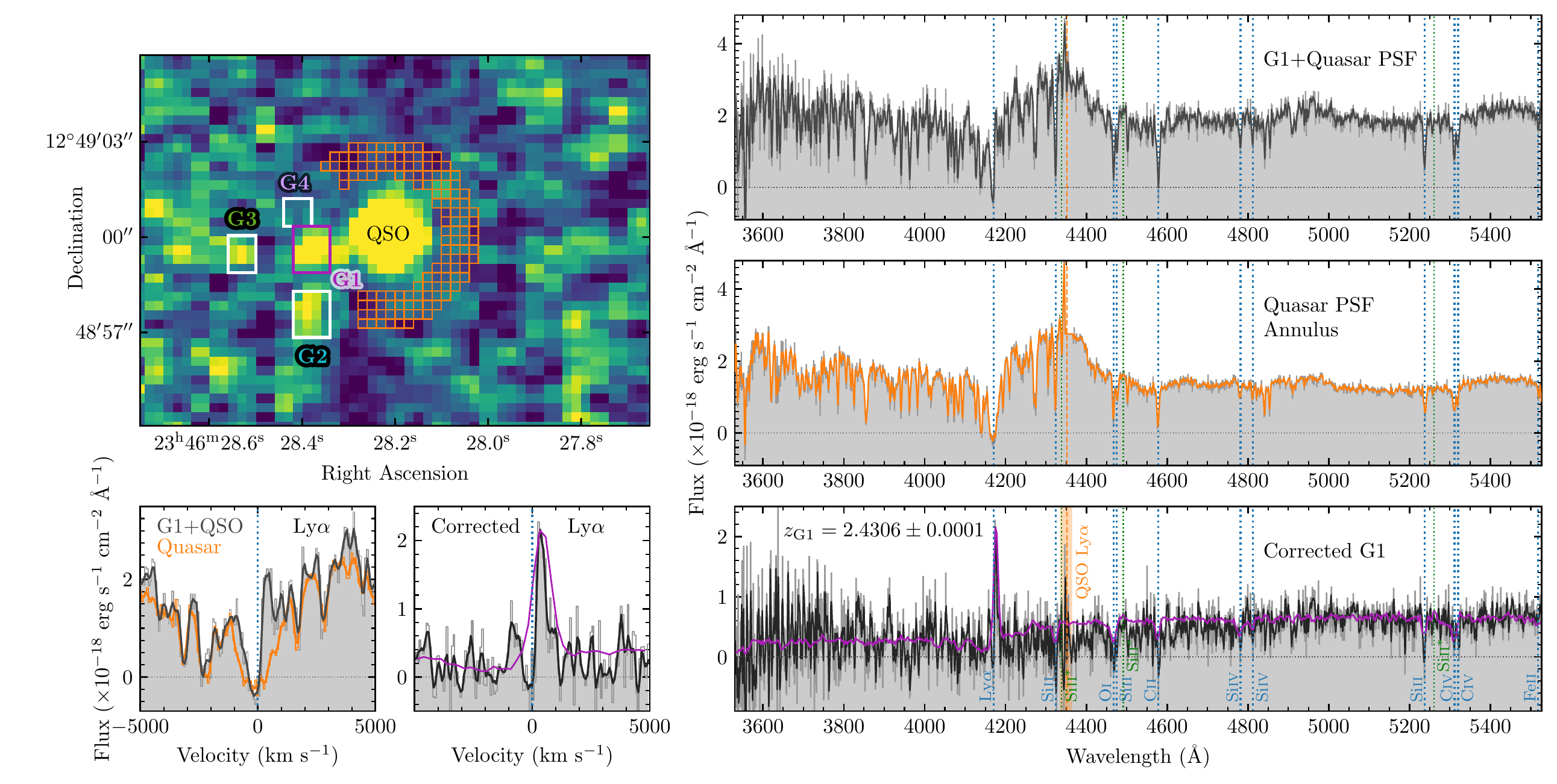}
    \caption{Galaxy spectrum extraction. ({\it top left}) KCWI line map image centred on the quasar with a velocity window of $v=z_{\rm DLA}\pm940$~{\kms}, which aims to minimise the quasar continuum in the image but still cover the galaxy {\Lya} emission lines. Galaxies G1, G2, G3, and G4 are labelled, though G4 is not visible in the image. The spaxels used to extract the spectrum of the galaxy nearest to the quasar sightline, G1, are contained within the purple rectangle, whereas the white rectangles highlight the spaxels extracted for the other galaxies. Orange squares indicate the spaxels used to estimate the quasar PSF in an annulus with a radius equal to G1's impact parameter and is as wide as the purple rectangle. ({\it right}) Extracted G1 spectrum ({\it top}) before and ({\it bottom}) after the ({\it middle}) quasar PSF annulus is subtracted. The thick purple curve represents a template Lyman break galaxy (LBG) spectrum shifted to the redshift of G1. Typical absorption (blue) and emission (green) lines in the LBG spectrum are labelled and plotted as vertical lines. The broad emission peak and strong absorption at $\lambda\sim4350$~{\AA} represents excess {\Lya} emission at $z_{\rm qso}$ due to an extended {\Lya}-emitting halo around the quasar and has been partially masked to reduce oversubtraction. ({\it bottom left}) Zoom-ins on the spectral region around the G1 {\Lya} line for the spectra plotted on the right.}
    \label{fig:galspec}
\end{figure*}

\section{Results}
\label{sec:results}

Combining the KCWI spectroscopy, UVES quasar spectrum, and {\it HST} imaging, we surveyed the galaxies around J2346+124 to determine their redshifts and to connect them to their CGM absorption or lack thereof. For this work, we specifically focused on finding the host galaxies for the known DLA in this field. To do this we created a pseudo narrow-band image from the KCWI data at the wavelengths expected to give rise to galaxy {\Lya} emission at $z_{\rm DLA}$ (see Fig.~\ref{fig:galspec}). The KCWI spectrum for each region that peaked in flux in the narrow band image was then further investigated to determine if the flux peaks were {\Lya} emission or noise. We did not require that the emission corresponded to a galaxy visible in the {\it HST} imaging. However, given we have deep {\it HST} imaging, we also extracted KCWI spectra from every galaxy visible in the {\it HST} data within the KCWI footprint and searched for identifiable emission and absorption features across the entire spectrum. Further details of our survey methods as well as the galaxy catalogues for each field in our survey will be presented in a future paper (Nielsen {\etal}, in preparation). Here we focus only on a group of galaxies giving rise to ultra-strong {\MgII} absorption in a DLA at $z_{\rm DLA}=2.431$ that were identified by their {\Lya} emission or ISM absorption.

\subsection{DLA Host Galaxies in a Group Environment}
\label{sec:galaxies}

From our survey of the J2346+124 field, we identified three galaxies with {\Lya} emission at $z_{\rm DLA}$ (G1, G2, and G3), and a fourth galaxy that has no obvious emission in KCWI but does have weak ISM absorption lines (G4). The galaxies are located at an angular separation of $\theta=2\farcs7-4\farcs9$ from the quasar, where their right ascensions, declinations, and angular separations from the quasar are listed in Table~\ref{tab:DLAgals} and their IDs are marked on Fig.~\ref{fig:whitelight}. The bottom right panel of the figure also shows a zoom-in on the galaxies, which are located within $\sim 1\farcs0-3\farcs5$ of each other and so are likely interacting. Apparent magnitudes in the three observed {\it HST}/WFC3 bands (F110W, F140W, and F160W) are tabulated in Table~\ref{tab:HSTgalprops} for each galaxy. The {\it HST} imaging demonstrates that these four galaxies have similar sizes and colours, while G1 and G4 appear to have tidal features in the direction of the other group members. These properties all suggest the galaxies are located at a similar redshift. Furthermore, the KCWI linemap in Fig.~\ref{fig:galspec}, which is a psuedo-narrow band image centred on {\Lya} at $z_{\rm DLA}$ with a velocity window of $v=z_{\rm DLA}\pm940$~{\kms}, shows flux peaks at the locations of G1, G2, and G3, indicating the presence of {\Lya} emission. There is no obvious flux peak at the location of G4, suggesting that it is not a {\Lya} emitter.

\subsubsection{Quasar PSF Subtraction and Galaxy Spectra Extraction}
\label{sec:PSF}

Given the close on-the-sky proximity of the galaxies to the background quasar, and since the quasar has its own expansive {\Lya} halo (not discussed here), the KCWI spectra for the closer galaxies, G1, G2, and G4 are contaminated by the quasar's point spread function (PSF). The contamination is observed in the KCWI whitelight image in Fig.~\ref{fig:whitelight} (upper right), where these inner galaxies are blended with the quasar light (while its spectrum is not greatly contaminated, G3 is still not visible in the whitelight image due to low continuum levels). In the spectra, the contamination is seen as a galaxy spectrum with some of the quasar's continuum or only its {\Lya} emission superimposed. Fig.~\ref{fig:galspec} demonstrates this effect for galaxy G1, where the spaxels extracted are highlighted as a purple rectangle in the KCWI line map (top left). The original G1 spectrum obtained from within the purple rectangle has been smoothed using a boxcar filter with a kernel of three pixels and is plotted in the top right panel of Fig.~\ref{fig:galspec}. This spectrum clearly has broad emission features that are the result of the quasar (see the middle right panel); notably the {\Lya} emission peak at $z_{\rm qso}$ is prominent redward of the expected location for {\Lya} emission at $z_{\rm DLA}$. We have partially masked this feature to reduce its prominence in the spectrum, but the orange vertical dashed line indicates the significant quasar {\Lya} emission contamination ($z_{\rm qso}=2.573$). Other broad emission features at wavelengths $\sim4950$~{\AA} and $\sim5450$~{\AA} correspond to {\SiIV} and {\CIV} at $z_{\rm qso}$. There are also strong absorption features labelled with the vertical blue dotted lines, which are the result of both the absorption in the surrounding CGM from the quasar light (i.e. metal lines associated with the DLA feature) as well as absorption due to the ISM of G1 itself.

\begin{figure*}
    \centering
    \includegraphics[width=\linewidth]{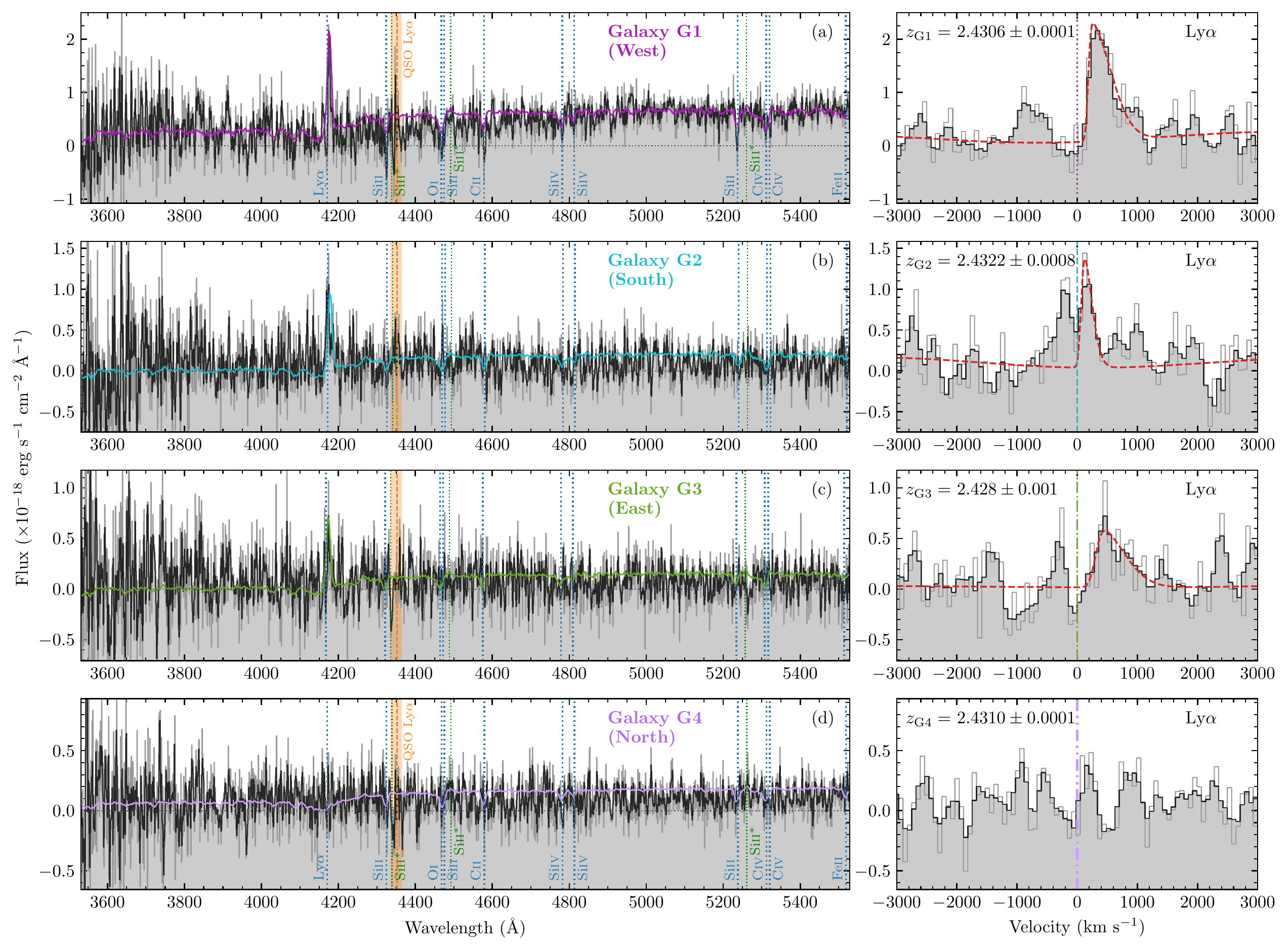}
    \caption{({\it left}) Spectra and ({\it right}) {\Lya} profiles for galaxies (a) G1, (b) G2, (c) G3, and (d) G4. Raw data and boxcar-smoothed data are plotted as the thin and thick black lines, respectively, with a LBG template overlaid in purple, cyan, green, or pink at the redshift of each galaxy. Vertical lines indicate typical absorption (blue) and emission (green) line features in the LBG templates. Orange vertical lines indicate contaminating {\Lya} from the background quasar at similar impact parameters (see Fig.~\ref{fig:galspec}). Red dashed curves in the right panels show a skewed Gaussian fit to the red peak of {\Lya} and galaxies G1 and G2 appear to have a double peak morphology in {\Lya}, however we use only the red peak in our redshift measurements. G3 may have a weak blue peak that is on the order of the noise level. G4 does not have any obvious {\Lya} emission above the level of the noise, although the peak at $v\sim200$~{\kms} could potentially be very weak {\Lya}. Vertical dashed lines in these panels show {\Lya} at the redshift labelled in each panel, which is used to set the velocity zero-point in each case.}
    \label{fig:galaxies}
\end{figure*}

To correct the galaxy spectra for the quasar PSF contamination, ideally we would model the PSF of another point source in the field using methods such as the empirical modeling in {\sc CWITools} \citep{cwitools} to account for any asymmetries in the PSF. However, we do not cover other point sources with our KCWI data so we must model the quasar itself. The empirical modeling method does not work well when continuum sources, e.g. galaxies, are blended with the PSF to be removed because the continuum levels of these objects we are attempting to isolate are subtracted with the quasar's PSF. Therefore we must assume the quasar PSF is axisymmetric and only model it at azimuthal angles where the galaxies are not present. This is a reasonable assumption because the quasar is a point source that is unresolved at every wavelength, it likely outshines its host galaxy (i.e. any asymmetries due to the quasar host galaxy are dwarfed by the quasar itself), and the quasar host galaxy would have a size that is on the order of the KCWI spaxel sizes and would be smaller than the seeing ($\sim0\farcs7-1\farcs0$; i.e. it is also unresolved). \citet{ychen21} demonstrated that the KCWI PSF is axisymmetric by comparing the PSFs of galaxies at $z=2-3$ in KCWI data with their corresponding PSFs in {\it HST} data convolved with the KCWI resolution.

Given this, we created a quasar PSF spectrum for each galaxy, which was extracted from spaxels in an annulus with the same range of galaxy--quasar angular separations as the spaxels extracted for each galaxy. This method is similar to the one employed by \citet{zabl21} to remove the quasar PSF at distances greater than $1\farcs4$ to minimize the contributions of continuum sources such as other galaxies. It has the benefit that we do not make an assumption for the profile shape of the quasar's PSF (i.e. Gaussian or Moffat), which can leave significant residuals in the PSF-subtracted image. Furthermore, full modeling of the quasar's PSF is not necessary for this field since previous work has already demonstrated that there are no host galaxies located on top of the quasar \citep{wang2015}. One drawback to this method is its assumption of complete axisymmetry; the whitelight image in Fig.~\ref{fig:whitelight} shows that the quasar has residual scattered light even after our custom reduction (e.g. Appendix~\ref{app:gradient}) and faint diffraction spikes, which are averaged over in the annuli. These features are not present in the {\Lya} linemap in Fig.~\ref{fig:galspec}, so they are unlikely to impact our {\Lya} emission identifications. We tested removing these diffraction spikes by rotating the field around the quasar by 180~deg and subtracting this from the original cube. We did not find a significant difference in measurements involving galaxy continuum levels with this method compared to our annulus method.

The quasar PSF annulus spaxels corresponding to G1 are highlighted as orange squares in Fig.~\ref{fig:galspec} (top left). Given our axisymmetry assumption, these annulus spaxels are assumed to be consistent with the quasar PSF within spaxels extracted for G1 (purple rectangle). We measured the median spectrum from these annulus spaxels and scaled the spectrum to the number of galaxy spaxels extracted (Fig.~\ref{fig:galspec} middle right) then subtracted the quasar's PSF spectrum from the raw galaxy spectrum. The resulting corrected G1 spectrum is shown in the bottom right panel of Fig.~\ref{fig:galspec}, where the thick black curve is G1's spectrum smoothed with a boxcar kernel of three pixels. We also plot a template Lyman break galaxy (LBG) spectrum \citep[purple; e.g.][]{shapley03} and vertical dotted lines highlighting absorption (blue) and emission (green) features for reference. The {\Lya} profiles for the galaxy and quasar annulus are also presented in the bottom left of Fig.~\ref{fig:galspec}. The DLA absorption is visible in the quasar annulus spectrum even at this large distance from the quasar. The {\Lya} emission from G1 presents as excess emission above the DLA in the quasar annulus spectrum that remains after the PSF subtraction. This quasar PSF subtraction was applied to G1, G2, and G4 with the appropriate quasar annulus for their radii. The quasar's PSF is reduced to background levels at radii corresponding to G3 so the PSF subtraction was not applied to G3. G4 has no obvious {\Lya} emission (see the upper left panel of Fig.~\ref{fig:galspec}) but we extracted a spectrum from the spaxels corresponding to its location in the {\it HST} imaging.

The resulting galaxy spectra from KCWI for G1, G2, G3, and G4 (quasar PSF subtracted if necessary) are plotted in Fig.~\ref{fig:galaxies}, where the left panels show the full spectrum and the right panels highlight the {\Lya} emission for each galaxy with a velocity window of $z=z_{\rm gal}\pm3000$~{\kms}. In each panel the spectrum is plotted as a thin black line, while the thick black lines are the spectrum smoothed using a boxcar filter with a kernel of three pixels. Template LBG spectra, which are good matches to the continuum shapes in the data, are plotted for reference in each spectrum as thick coloured lines. We observe the red peak of {\Lya} emission for G1, G2, and G3, with the blue peak also present in G1 and G2 (G3 may have a blue peak, but it is on the level of the noise). G4 does not have any obvious {\Lya} emission above the level of the noise, but does have weak ISM absorption lines in its spectrum. None of the four galaxies has prominent absorption or emission features that suggest they are located at a redshift different from roughly $z_{\rm DLA}$. Our methods for measuring the galaxy properties and the derived values for each galaxy are discussed in the following sections.

\subsubsection{Galaxy G1 (West)}
\label{sec:G1}

G1 is located $\theta=2\farcs7$ directly east of the quasar sightline and is the brightest member of the group in all of the {\it HST} bands, KCWI, and in {\Lya} emission. It is the westernmost galaxy in the group, with no {\Lya} emitters on top of the quasar. A previous survey of the field did not survey far enough to cover G1 or the rest of the compact group (i.e. no {\Halpha}-emitting galaxies were located within $\sim1\farcs5$ or 12.5~kpc of the quasar to SFR limits ranging ${\rm SFR}>4.4-11$~M$_{\odot}$~yr$^{-1}$; \citealp{wang2015}). G1 has both {\Lya} emission and ISM absorption lines, which are labelled in the KCWI spectrum of Fig.~\ref{fig:galaxies} (top). We measured a redshift using several ISM absorption lines covered in the KCWI data. These resulted in redshifts of $z_{{\small \SiII}~\lambda 1260}=2.4316\pm0.0002$ and $z_{{\small \SiII}~\lambda 1526}=2.4296\pm0.0002$, with an average of $z_{{\small \SiII}}=2.4306\pm0.0001$. 

We also measured the redshift from {\Lya}, which is typically observed as an asymmetric red peak with a tail to longer wavelengths for galaxies with a large enough escape fraction of {\Lya} photons, $f_{\rm esc}$. A blue peak is more rare. This is because {\Lya} is subject to resonant scattering by the galaxy itself as well as its surrounding CGM, where the observed emission profile depends on the kinematics and amount of gas that the photons must pass through. Previous work has found that the redshift obtained from the line is often offset from the redshift obtained from ISM absorption lines on the order of a few hundred {\kms} \citep[e.g.][]{zheng02, shapley03, trainor15}. Given this, we used the methods of \citet{verhamme18}, who estimate the velocity offset between the redshift obtained with the red peak of {\Lya} and the true galaxy redshift, $v_{\rm red}$, by using the full width at half maximum (FWHM) of the red peak. They also provide a method that estimates the velocity offset using the velocity separation between the red and blue {\Lya} peaks where both are observed, but we do not use it even where a blue peak is present in order to be consistent between galaxies. From Fig.~\ref{fig:galaxies} (right) the {\Lya} profile for G1 clearly has a red peak and a weaker blue peak. 

To measure $z_{\scriptsize \Lya}$, we fit a skewed Gaussian to the red peak and plot this as a red dashed line in the upper right panel of Fig.~\ref{fig:galaxies}. We measured the wavelength of peak flux in the fit and converted it to a redshift. This resulted in a value of $z_{\scriptsize \Lya}=2.4337\pm0.0001$, which is offset from the value measured from the {\SiII} ISM absorption lines by $180-355$~{\kms}. The FWHM of the red peak is 430~{\kms} and, using the \citeauthor{verhamme18} relation $v_{\rm red}=0.9\times{\rm FWHM}-34$~{\kms}, corresponds to a velocity offset of $v_{\rm red}=353$~{\kms} from the true redshift. This FWHM-derived velocity offset is equal to the offset between $z_{\scriptsize \Lya}$ and $z_{\SiII~\lambda1526}$, which confirms the ability for the \citeauthor{verhamme18} method to correctly measure the redshift from {\Lya}. Applying this velocity shift to the observed redshift via $z_{{\scriptsize \Lya},{\rm corr}} = z_{\scriptsize \Lya} - (v_{\rm red}/c)(\lambda_{\rm obs}/1215.67)$, we obtained $z_{{\scriptsize \Lya},{\rm corr}}=2.4296\pm0.0008$. This value is consistent within uncertainties with the values measured with ISM absorption lines. 

While our three redshift estimates have provided values that are all within uncertainties, we adopt an average redshift between the ISM {\SiII} lines since they have the least systematic error. Thus G1 has an adopted galaxy redshift of $z_{\rm G1}=2.4306\pm0.0001$, which is labelled on the upper right panel of Fig.~\ref{fig:galaxies} and is used to set the velocity zero point of the panel. This value and the derived galaxy properties are tabulated in Tables~\ref{tab:HSTgalprops} and \ref{tab:KCWIgalprops}. With this redshift, G1 has an impact parameter of $D=21.5$~kpc. We converted $m_{\rm F160W}$ to an absolute $B$-band magnitude via the methods described in \citet{magiicat1} and report a value corresponding to an Scd galaxy SED. Uncertainties due to the selection of SED type are less than $\pm0.05$ magnitudes, which represents the range in values allowed for Sbc and Im SED types. We also converted the absolute magnitude, $M_B=-22.7$, to a $B$-band luminosity using the relation between $M^{\ast}$ and redshift from \citet{gabasch04}, who measure $B$-band luminosity functions over $0.5<z<5.0$. G1 is roughly an $L^{\ast}$ galaxy with $L_B/L_B^{\ast}=1.66$. It has an observed colour of ${\rm F110W}-{\rm F160W}=0.5$, which is comparable to the other galaxies in the group. For comparison to the literature, we measured the UV absolute magnitude, $M_{\rm UV}$, by integrating between rest-frame wavelengths $1400-1600$~{\AA} in the KCWI spectrum. For G1, we found $M_{\rm UV}=-21.9$.

We estimated galaxy morphologies from the {\it HST}/WFC3 F160W band (see Fig.~\ref{fig:whitelight}, bottom right), which was chosen over the other available bands because it most accurately traces the bulk of the starlight in the galaxy since it is the reddest band of the three (roughly rest-frame $B$-band). In contrast, F110W traces the redder end of near-UV wavelengths that are dominated by the youngest stars, which tend to be located in star-forming clumps \citep[e.g.][]{guo15, shibuya16} and do not necessarily trace the majority of the mass distribution. Briefly, the time- and position-dependent point spread functions for the galaxies in the field were modelled using Tiny Tim \citep{tinytim} and used in the galaxy modelling. We quantified morphological properties by fitting a single-component model using {\sc GIM2D} \citep{simard02}, where the component has a S\'{e}rsic profile ($0.2\leq n\leq4.0$). A single-component model was used in contrast to the two-component model we have used previously \citep[e.g.][]{kacprzak15, nielsen20} because it is more simple (less degeneracy) and the constraints on these galaxy morphologies are not adequate enough to justify a second component given their close proximity to the quasar and each other (i.e. it is highly likely that these galaxies have disturbed morphologies from interactions). The quasar's sightline through the foreground CGM was then compared to the galaxy's on-the-sky orientation (i.e. its projected major axis). An azimuthal angle of $\Phi=0^{\circ}$ represents a sightline along the projected major axis of the galaxy, whereas $\Phi=90^{\circ}$ corresponds to a sightline along the projected galaxy minor axis. Galaxy inclinations were defined such that $i=0^{\circ}$ is a face-on galaxy and $i=90^{\circ}$ is an edge-on galaxy.

Using this method, galaxy G1 is best modelled with $i=66^{\circ}$ and $\Phi=17^{\circ}$, classifying it as an intermediately-inclined galaxy with the quasar probing closest to the projected major axis and a half-light radius of $R_{1/2}=6.6$~kpc (Table~\ref{tab:HSTgalprops}). These morphologies still have considerable uncertainties that are not reflected in the table since the galaxies are quite small in the {\it HST} imaging, making the PSF modelling important, and they are near the quasar, which contaminates their profiles. While we report them here, we do not place much weight on the morphologies for the interpretation of the system. Additionally, the galaxy morphology is less likely to be useful in group environments than for isolated galaxies due to the additional complexity added to the CGM gas flows from galaxy interactions. These interactions may also result in tidal features around the galaxies that could influence the morphology modelling. In fact, G1 appears to have material in a tidal stream in the southeast direction in all of the images and is likely resulting in the major axis measurement, although the quasar complicates confirmation of this feature.

\begin{table*}
	\centering
	\caption{Properties of the DLA Host Galaxies from the {\it HST} Imaging}
	\label{tab:HSTgalprops}
	\begin{threeparttable}
	\begin{tabular}{lcccccccccc}
		\hline
        Galaxy     &  $z_{\rm gal}$    &  $D$   &  $m_{\rm F110W}$ & $m_{\rm F140W}$ & $m_{\rm F160W}$ & $M_{B}$ & $L_B/L_B^{\ast}$ & $i$              & $\Phi$           & $R_{1/2}$  \\
                   &                   &  (kpc) &  (AB)            & (AB)            & (AB)            & (AB)    &                  & (deg)            & (deg)            & (kpc)      \\
        \hline                                                                                                                                                        
        G1 (West)  & $2.4306\pm0.0001$ &  21.5  & $22.8\pm0.2$     & $22.6\pm0.1$    & $22.3\pm0.2$    & $-22.7$ & $1.66$           & $66^{+6}_{-7}$   & $17^{+10}_{-8}$  &  $6.6$ \\[3pt]
        G2 (South) & $2.4322\pm0.0008$ &  30.6  & $24.2\pm0.4$     & $24.2\pm0.3$    & $25.1\pm0.7$    & $-19.9$ & $0.12$           & $58^{+8}_{-12}$  & $14^{+14}_{-14}$ &  $2.9$ \\[3pt]
        G3 (East)  & $2.428\pm0.001$   &  39.6  & $25.5\pm0.5$     & $25.1\pm0.4$    & $24.9\pm0.7$    & $-20.1$ & $0.15$           & $30^{+11}_{-12}$ & $12^{+26}_{-12}$ &  $7.2$ \\[3pt]
        G4 (North) & $2.4310\pm0.0001$ &  26.1  & $24.5\pm0.7$     & $23.7\pm0.2$    & $24.1\pm0.4$    & $-20.9$ & $0.33$           & $65^{+5}_{-5}$   & $63^{+6}_{-5}$   & $12.1$ \\
		\hline
	\end{tabular}
    \end{threeparttable}
\end{table*}

\begin{table*}
    \centering
    \caption{Properties of the DLA Host Galaxies from KCWI Spectroscopy}
    \begin{threeparttable}
    \begin{tabular}{lcccccccc}
        \hline
        Galaxy     &  $z_{\rm gal}$    & $D$   & $M_{\rm UV}$  & $\log L_{\scriptsize \Lya}$ & $W_r({\Lya})$ & SFR$_{\scriptsize \Lya}$\tnote{a} & SFR$_{\rm FUV}$\tnote{b} & $\Sigma_{\rm SFR}$\tnote{b}   \\
                   &                   & (kpc) & (AB)          &                             & ({\AA})       & (M$_{\odot}$~yr$^{-1}$)           & (M$_{\odot}$~yr$^{-1}$)  & (M$_{\odot}$~yr$^{-1}$~kpc$^{-2}$)\\
        \hline
        G1 (West)  & $2.4306\pm0.0001$ &  21.5 & $-21.9$       & 41.95                       & 70.9          & $>1.3$                            & 11.6                     & 0.085    \\
        G2 (South) & $2.4322\pm0.0008$ &  30.6 & $-19.8$       & 41.65                       & 68.1          & $>0.7$                            & 1.7                      & 0.063    \\
        G3 (East)  & $2.428\pm0.001$   &  39.6 & $-19.6$       & 41.34                       & 34.3          & $>0.7$                            & 1.4                      & 0.008    \\
        G4 (North) & $2.4310\pm0.0001$ &  26.1 & $-20.1$       & $<40.46$                    & $\cdots$      & $\cdots$                          & 2.0                      & 0.004    \\
        \hline
    \end{tabular}
    \begin{tablenotes}
                \item[a] Assuming $f_{\rm esc,LyC}=0.0$ as a conservative lower limit and a Chabrier IMF \citep[e.g.][]{sobral19} converted to a Kroupa IMF.
                \item[b] Assuming a Kroupa IMF and a conservative median $E(B-V)_{\rm neb}=0.26$ for the KBSS sample \citep{trainor19} with $A_{\rm FUV}=7.6 E(B-V)$ \citep{hao11}. Using an upper interquartile value of $E(B-V)_{\rm neb}=0.47$ from KBSS results in ${\rm SFR}_{\rm FUV}$ values that are $\sim 4.3$ times larger than listed here (e.g. 50.4, 7.2, 5.9, and 8.7~M$_{\odot}$~yr$^{-1}$ for G1$-$G4, respectively).
    \end{tablenotes}
    \end{threeparttable}
    \label{tab:KCWIgalprops}
\end{table*}

With the KCWI data, we estimated a star formation rate, SFR, using the {\Lya} emission line and the methods described in \citet{sobral19}. In short, the ${\rm SFR}_{\scriptsize \Lya}$ can be determined from the rest {\Lya} equivalent width, $W_r(\Lya)$, and the {\Lya} luminosity, $L_{\scriptsize \Lya}$. Assuming an escape fraction for {\Lya} photons, $f_{\rm esc, LyC}$, this results in a SFR that is comparable to dust-corrected {\Halpha} SFRs for $z=0-2.6$. We used the relation between these observables and ${\rm SFR}_{\scriptsize \Lya}$ for a Chabrier initial mass function (IMF) and $f_{\rm esc, LyC}=0$, then converted to a Kroupa IMF \citep[by multiplying by 1.06;][]{speagle14}. Given the uncertainty in the escape fraction and the dust extinction within G1, we report the SFR as a lower limit. For G1, we observe both the blue and red {\Lya} peaks with $W_r(\Lya)=70.9$~{\AA} and luminosity $\log L_{\scriptsize \Lya}=41.95$, which are listed in Table~\ref{tab:KCWIgalprops}. From the \citeauthor{sobral19} relation we estimate ${\rm SFR}_{\scriptsize \Lya}>1.3$~M$_{\odot}$~yr$^{-1}$. If we instead use $f_{\rm esc, LyC}=0.15$, which is the upper range typical for {\Lya} emitters \citep{matthee17, verhamme17}, we estimate ${\rm SFR}_{\scriptsize \Lya}>1.6$~M$_{\odot}$~yr$^{-1}$.

Given the uncertainties in the SFR derived from {\Lya}, we also calculated ${\rm SFR}_{\rm FUV}$ with the relation in \citet{hao11} for a Kroupa IMF with solar metallicity and 100~Myr age to compare to the ${\rm SFR}_{\scriptsize \Lya}$. \citeauthor{hao11} also provide a relation to estimate the dust attenuation in the FUV band, $A_{\rm FUV}$, using the NUV band. However, the KCWI spectrum covers only the bluer half of the rest-frame FUV band while the bluest {\it HST} filter, F110W, only covers the redder half of the rest-frame NUV band. This does not provide an accurate measure of the dust attenuation in this system. Instead, we used the median $E(B-V)_{\rm neb}$ from the KBSS, $E(B-V)_{\rm neb}=0.26$ \citep{strom17, trainor19, theios19}, which was measured from the Balmer decrement, and the relation $A_{\rm FUV}=7.6 E(B-V)$ from \citet{hao11} to estimate the typical extinction expected for $z\sim2$ galaxies. We then corrected the $m_{\rm UV}$ measured from KCWI between rest-frame wavelengths $1400-1600$~{\AA} using this attenuation and applied the \citeauthor{hao11} SFR relation. G1 has ${\rm SFR}_{\rm FUV}=11.6$~M$_{\odot}$~yr$^{-1}$ assuming the median KBSS $E(B-V)_{\rm neb}$. If we instead use the interquartile $E(B-V)_{\rm neb}$ values reported for KBSS ($0.06-0.47$), the ${\rm SFR}_{\rm FUV}$ ranges from $2.9-50.4$~M$_{\odot}$~yr$^{-1}$. The lower value of this range is similar to the SFR measured with {\Lya}, but the upper value suggests that this galaxy could be undergoing significant star formation activity, which would be expected for galaxy--galaxy interactions. We adopt the SFR estimated from the median dust extinction to be conservative, which roughly places G1 on the star formation main sequence at this redshift \citep[e.g.][]{speagle14}. 

Using the half-light radius estimated from the {\it HST} imaging and the median dust attenuation, we find a star formation rate surface density of $\Sigma_{\rm SFR}=0.085$~M$_{\odot}$~yr$^{-1}$~kpc$^{-2}$. This value is just below the threshold typically found to drive outflows at lower redshifts \citep[$0.1$~M$_{\odot}$~yr$^{-1}$~kpc$^{-2}$, e.g.][]{heckman02, heckman15}. However, the upper interquartile SFR value assuming more dust attenuation has $\Sigma_{\rm SFR}=0.368$~M$_{\odot}$~yr$^{-1}$~kpc$^{-2}$, which is larger than the threshold. G1 may be driving an outflow that could result in metal-rich absorption in the quasar spectrum. It could also give rise to metal-poor accretion signatures since the galaxy is probed along its major axis and it must be accreting fresh material for its star formation activity.


\subsubsection{Galaxy G2 (South)}

G2 is located $\theta=3\farcs8$ southeast of the quasar (Table~\ref{tab:DLAgals}) and has both red and blue peaks in {\Lya} emission, but no identifiable ISM absorption lines due to the low continuum level in the KCWI data (Fig.~\ref{fig:galaxies}). Its redshift is $z_{\rm G2}=2.4322\pm0.0008$, which was measured from the red {\Lya} emission peak following the methods described for G1 in Section~\ref{sec:G1} (Table~\ref{tab:KCWIgalprops}). The galaxy then has an impact parameter of $D=30.6$~kpc making it the third-nearest galaxy to the quasar. G2 is the bluest galaxy of the group with an observed F110W-F160W colour of $-0.9$, which could either be a result of a higher star formation rate or simply less dust extinction than the other galaxies. The latter may be more correct since we observe both the red and blue peaks of {\Lya} where typically only a red peak is observed, and the blue peak has a similar strength as the red peak. Using both of these peaks, we estimated ${\rm SFR}_{\scriptsize \Lya}>0.7$~M$_{\odot}$~yr$^{-1}$ for $f_{\rm esc, LyC}=0$ (${\rm SFR}_{\scriptsize \Lya}>0.9$~M$_{\odot}$~yr$^{-1}$ for $f_{\rm esc, LyC}=0.15$). Assuming the median KBSS $E(B-V)_{\rm neb}$, the ${\rm SFR}_{\rm FUV}=1.7$~M$_{\odot}$~yr$^{-1}$ is also low, but could be as large as 7.2~M$_{\odot}$~yr$^{-1}$ if we assume the upper interquartile dust extinction value. Using the median FUV SFR, we find $\Sigma_{\rm SFR}=0.063$~M$_{\odot}$~yr$^{-1}$~kpc$^{-2}$, which is below the threshold. However, the upper interquartile dust extinction FUV SFR gives $\Sigma_{\rm SFR}=0.274$~M$_{\odot}$~yr$^{-1}$~kpc$^{-2}$, which could suggest that the galaxy is star-forming enough to drive outflows. G2 is a sub-$L^{\ast}$ galaxy with absolute magnitude $M_B=-19.9$ and luminosity $L_B/L_B^{\ast}=0.12$ (Table~\ref{tab:HSTgalprops}) and a UV absolute magnitude $M_{\rm UV}=-19.8$ (Table~\ref{tab:KCWIgalprops}). From the {\sc GIM2D} modeling, we find that G2 is best modelled with a half-light radius of $R_{1/2}=2.9$~kpc, an intermediate inclination, $i=58^{\circ}$, and with the quasar sightline probing its projected major axis, $\Phi=14^{\circ}$. Similar to G1, G2 may give rise to both metal-rich outflowing and metal-poor accretion signatures.

\subsubsection{Galaxy G3 (East)}

Galaxy G3 is the furthest of the group galaxies from the quasar with $\theta=4\farcs9$ directly east of the quasar (Table~\ref{tab:DLAgals}). It has a red {\Lya} peak, with a possible blue peak that is on the level of the noise. The red peak gives a corrected redshift of $z_{\rm G3}=2.428\pm0.001$ following the methods described for G1 (Table~\ref{tab:KCWIgalprops}), but there are no identifiable ISM absorption features in the spectrum due to the low continuum level (Fig.~\ref{fig:galaxies}). G3 is then at an impact parameter of $D=39.6$~kpc and is a sub-$L^{\ast}$ galaxy with $M_{\rm UV}=-19.6$, $M_B=-20.1$, and $L_B/L_B^{\ast}=0.15$ (Table~\ref{tab:HSTgalprops}). It has the reddest observed F110W$-$F160W colour out of the group (0.6) and a low SFR with ${\rm SFR}_{\scriptsize \Lya}>0.7$~M$_{\odot}$~yr$^{-1}$, measured using the \citet{sobral19} relation with only the red {\Lya} peak and $f_{\rm esc, LyC}=0$ (${\rm SFR}_{\scriptsize \Lya}>0.8$ for $f_{\rm esc, LyC}=0.15$). The ${\rm SFR}_{\rm FUV}=1.4$~M$_{\odot}$~yr$^{-1}$ is also low, but could be as large as 5.9~M$_{\odot}$~yr$^{-1}$ assuming more dust extinction. The SFR surface density of $\Sigma_{\rm SFR}=0.008$~M$_{\odot}$~yr$^{-1}$~kpc$^{-2}$ is too low to drive significant outflows, even assuming more dust extinction ($\Sigma_{\rm SFR}=0.037$~M$_{\odot}$~yr$^{-1}$~kpc$^{-2}$). G3 is best modelled as an intermediately-inclined, $i=30^{\circ}$, galaxy with the quasar probing its projected major axis, $\Phi=12^{\circ}$, with a half-light radius of $R_{1/2}=7.2$~kpc. Given its properties, G3 would likely give rise to accretion signatures in the quasar sightline, but would not have recently contributed outflowing material.

\subsubsection{Galaxy G4 (North)}

The fourth galaxy in this group, G4, is located $\theta=3\farcs3$ east and slightly north of the quasar sightline. It is not directly observed in the KCWI line map in Fig.~\ref{fig:galspec} and there is no obvious {\Lya} emission in its spectrum (Fig.~\ref{fig:galaxies}d), which suggests it is either not a {\Lya} emitter or is at a different redshift than this group. G4 is of similar size, luminosity, and observed colour (${\rm F110W}-{\rm F160W}=0.4$) as the other three galaxies in the group so it is likely that it is also associated, and the F160W image appears to show it has tidal features in the directions of G3 and G1/G2. Close inspection of the galaxy's spectrum shows weak ISM lines, particularly in the {\SiIV} doublet, which give an average redshift of $z_{\rm G4}=2.4310\pm0.0001$. We cannot reliably estimate a redshift from {\Lya} as was done for the other three galaxies since any potential emission is on the level of the noise in the spectrum (e.g. the peak at $\sim20$~{\kms} in Fig.~\ref{fig:galaxies}(d)). Given this, G4 has an impact parameter of $D=26.1$~kpc. It is another sub-$L^{\ast}$ galaxy in the group, with $M_B=-20.9$, $L_B/L_B^{\ast}=0.33$, and $M_{\rm UV}=-20.1$ but is likely more massive than G2 and G3. Since we do not have {\Lya} emission for this galaxy and the {\Lya}-derived SFR is so uncertain, we do not report ${\rm SFR}_{\scriptsize \Lya}$. The FUV SFR is more reliable since it depends on the broadband magnitudes, where we estimate ${\rm SFR}_{\rm FUV}=2.0$~M$_{\odot}$~yr$^{-1}$ (could be as high as $8.7~$M$_{\odot}$~yr$^{-1}$ with $E(B-V)_{\rm neb}=0.47$). This gives a SFR surface density of $\Sigma_{\rm SFR}=0.004$M$_{\odot}$~yr$^{-1}$~kpc$^{-2}$, which is too low to drive outflows, even if we assume more dust ($\Sigma_{\rm SFR}=0.019$~M$_{\odot}$~yr$^{-1}$~kpc$^{-2}$). The modelling of the galaxy in the {\it HST} imaging suggests that G4 has an intermediate inclination, $i=65^{\circ}$, with an intermediate azimuthal angle, $\Phi=63^{\circ}$. It is estimated to be the largest galaxy in the group, with a half-light radius of $R_{1/2}=12.1$~kpc. However, the morphology of G4 is the least trustworthy in the group given its proximity to the quasar sightline and a diffraction spike in the {\it HST} image, as well as the possibility of multiple tidal features.

\subsubsection{The Compact Group Environment}

The DLA host galaxy environment thus consists of one $L^{\ast}$ galaxy and three sub-$L^{\ast}$ galaxies. They are separated by $8-28$~kpc from each other and there are potentially tidal structures present in both G1 and G4. There are no other galaxies identified at the redshift of this group within the KCWI FOV. These characteristics classify these galaxies as a compact group \citep[e.g.][]{hickson97, mcconnachie09}. 
The group geometric centre has an angular separation of $\theta=3\farcs4$ from the quasar, corresponding to $D=27.7$~kpc. If we instead weight by luminosity (as a proxy for mass), then the group luminosity weighted centre has $\theta=2\farcs9$ and $D=23.4$~kpc. 

All four galaxies are star-forming, although G1 is more active than G2, G3, and G4 by at least an order of magnitude. Both G1 and G2 have star formation rate surface densities above the fiducial threshold to drive outflows if we assume the upper range of dust extinction, whereas G3 and G4 are lower by over an order of magnitude and likely not driving outflows. This suggests that at least some subset of the galaxies are actively polluting or have recently polluted their CGM and IGrM with the metals observed in the quasar spectrum. It is not clear from the galaxy morphologies and small impact parameters that one galaxy should dominate the quasar absorption over the others. All galaxy inclinations are intermediate values but most azimuthal angles are major axis sightlines, which is the most likely orientation expected to trace accretion and tidal stripping. While none of the galaxies have minor axis sightlines or face-on inclinations, we expect the absorption to also trace outflows, especially if outflows are isotropic at this epoch \citep[e.g.][]{steidel10, nelson19} or if the galaxies are rapidly changing their locations about the group, both of which would make azimuthal angle measurements less important for gas interpretations. These features suggest that the physical origin of the absorbing gas in this compact group could be a combination of several different mechanisms including outflows, accretion, and tidal stripping.

\subsection{Cloud-by-Cloud Multiphase Bayesian Modelling}
\label{sec:cmbm}

We apply a new approach to characterise the detailed kinematic structure and physical conditions in the multiphase absorbing gas for this DLA: the cloud-by-cloud multiphase Bayesian ionisation modelling (CMBM) method. The CMBM method is fully detailed and tested in \citet{sameer2021} for several weak {\MgII} systems and updated in \citet{sameerleo} for multiple sightlines through a dense group environment (see the flowcharts in Fig.~2 in both works). This method differs from previous modelling in two ways. We characterise the physical conditions in (1) every cloud along the sightline as opposed to obtaining a single total column density for each ion for comparison to {\sc Cloudy} model grids and (2) for multiple ionisation phases, typically including the low and intermediate phases and, where necessary, a collisionally ionised phase, in contrast to assuming a single phase regardless of the ions included in the analysis. We summarise the method here. 

CMBM uses {\sc Cloudy} \citep{cloudy2017} to infer the physical conditions of the absorbing gas by synthesising absorption profiles given a set of photoionisation equilibrium models and comparing them to the observed absorption profiles. We adopt the \citet{ks19} model for extragalactic background radiation and assume a solar abundance pattern \citep{grevesse10}. {\sc Cloudy} model grids are generated for the ranges {\metallicity} $ \in[-3.0,1.5]$, $\log (n_{\rm H}/{\rm cm}^{-3})\in[-6.0,1.0]$, and $\log (N(\HI)/{\rm cm}^{-2})\in[11.0,21.0]$, with a step size of 0.1~dex. Each modelled cloud from {\sc Cloudy} is described by the parameters redshift, $z$, metallicity, {\metallicity}, hydrogen number density, {\hden}, {\HI} column density, {\colden}, and the non-thermal Doppler $b$-parameter, $b_{\rm nt}$. 

As a starting point for exploring the {\sc Cloudy} grids, initial Voigt profile (VP) component fits are obtained for constraining ions that are unsaturated and have high S/N in order to best determine the kinematic structure (i.e. the clouds that need characterising along the line-of-sight) and Doppler $b$-parameters. VP components are fit to the low ionisation lines first using {\HI}, {\MgII}, {\FeII}, or {\SiII} as the constraining ion, making the assumption of a single, low ionisation phase. {\HI} is only used as the constraining ion for components where metal lines are not present, while most low ionisation components were defined using {\SiII}. These constraining lines are used to define components, which can be adjusted in velocity, but which are applied to every transition, tying the kinematic structure across ions and transitions together. The column density, $b$-parameter, and redshift for each VP component is determined by fitting a Voigt-Hjerting function as implemented in the {\sc VoigtFit} package~\citep{krogager2018voigtfit, krogager18vfitsoftware}, and the best-fitting parameters are determined using the Nested sampling algorithm {\sc PyMultiNest}~\citep{pymultinest}. From these fits we use only the number of components and the component redshifts as initial positions which will be adjusted, and the Doppler parameters as the maximum possible value of $b_{\rm nt}$ in the prior distribution. We adopt a component structure with the least model complexity, i.e. the fewest number of components required to describe the absorption profile. 

Using this VP component structure as a starting point, synthetic absorption profiles with cloud parameters from the {\sc Cloudy} grids are generated, and compared to the observed profiles to obtain the best-fitting parameters and posterior distributions for each parameter and cloud (i.e. ``cloud-by-cloud''), again using {\sc PyMultiNest}. We adopt flat priors for {\metallicity}, {\hden}, {\colden}, and $b_{\rm nt}$. For $z$, a Gaussian prior centred on the initial component structure is adopted, with $\sigma$ equal to the error in $z$ from the initial VP fit. Where the model does not agree with the observations, the CMBM method adds additional clouds to the VP component structure to model the additional absorption (i.e. the gas is multiphase) and the comparison is conducted again. For example, if the model using only the low ionisation lines does not account for all of the absorption in higher ionisation lines, which indicates that a single phase does not fully describe the absorption, additional components are then fitted using an intermediate ionisation line such as {\SiIV} and {\CIV} as the constraining ion. Adding VP components to account for additional ionisation phases and kinematic structure, and then comparing the {\sc Cloudy} models to the observations is done iteratively until the models agree with the data. 

The results of the CMBM method are presented in Section~\ref{sec:absorption} for the VP component structure, Section~\ref{sec:metallicity} for the metallicity and ionisation conditions, and Appendix~\ref{app:metallicity} for the individual VP component models and physical properties. We did not use {\Lya} to constrain our modelling in this case because the location of the continuum of the DLA is quite uncertain and we can obtain accurate fits to the {\HI} using the rest of the Lyman series. This is the first time CMBM has been applied to such a complicated absorber, and the amount of detail it provides allows us to study how the photoionisation conditions change along the sightline and for different phases. It is especially important to understand these variations for interpreting the physical origin of the gas since several different physical structures could be located along the line-of-sight, especially in a compact group environment.

\subsection{Absorption System and Kinematics}
\label{sec:absorption}

\begin{figure*}
    \centering
    \includegraphics[width=\linewidth]{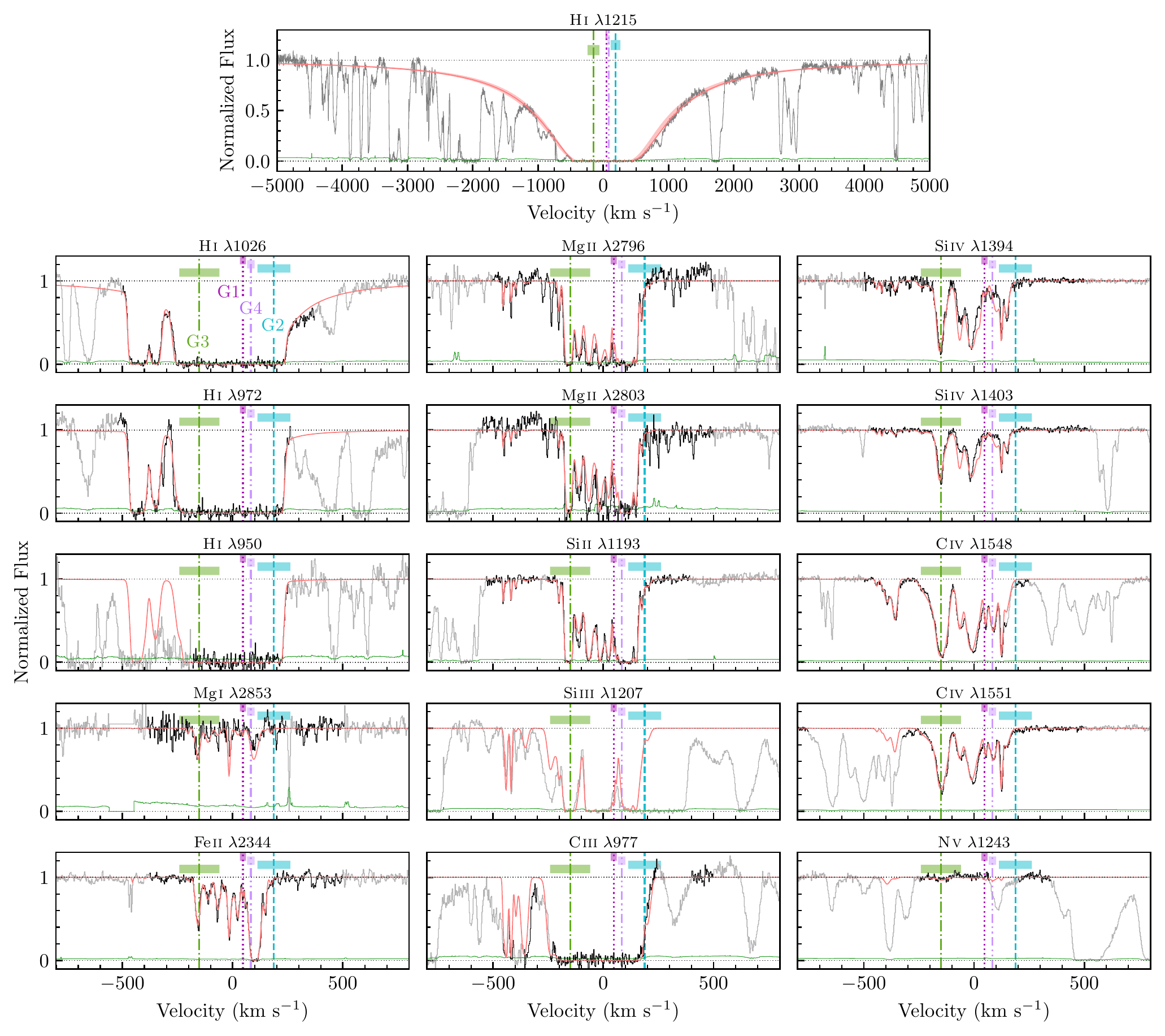}
    \caption{Absorption system at $z_{\rm abs}=2.43014$ in UVES, where we show a selection of lines that represent the various covered ionisation phases. Black lines represent the data, grey lines are the sections of the spectrum that were not considered in the CMBM analysis, and the green line is the error spectrum. The velocity zero-point corresponds to the optical depth-weighted median of {\MgII} absorption. The orange curve is the total fit from the metallicity modeling, accounting for multiple components and ionisation phases (see Section~\ref{sec:cmbm} and Appendix~\ref{app:metallicity}).~{\HI}~$\lambda1215$ has a wider velocity range to show the wings of the DLA and the orange shaded region demonstrates the uncertainty on the model. Vertical lines indicate the redshifts of the galaxies in the field, G1 (purple dotted), G2 (cyan dashed), G3 (green dot-dashed), and G4 (pink dot-dot-dashed). The shaded bars at the top of each panel show the redshift uncertainties of the galaxies and are stacked to represent their relative distance from the quasar, where G1 is closest to the quasar while G3 is furthest. The galaxy redshifts span the strongest metal line absorption kinematic spread, where G1, G2, and G4 cluster around the strongest components, but do not span the full range of {\HI} velocities, particularly the components with $v\sim-400$~{\kms}.}
    \label{fig:absorption}
\end{figure*}

The $z_{\rm abs}=2.43014$ absorption system is a DLA and was previously studied by several groups \citep[e.g.][and references therein]{dodorico2002, d-z2004, wang2015}. Fig.~\ref{fig:absorption} highlights a few of the transitions covered in our VLT/UVES spectrum, including {\HI}, {\MgI}, {\MgII}, {\FeII}, {\SiII}, {\SiIII}, {\SiIV}, {\CIII}, {\CIV}, and {\NV}. The selected transitions are those for which there are no significant blends (with the exception of {\SiIII}), are weak enough to show the kinematic structure, and span both the low and intermediate ionisation phases. Much of {\SiII}~$\lambda1193$ is saturated, however the weaker transitions are blended with other lines and so the kinematic structure is not as clear. While only the four strongest {\HI} transitions are plotted, we have coverage of the full Lyman series. We did not use the {\Lya} absorption as a constraint in our modelling due to uncertainties on the continuum fit but show the transition in the figure. The data are plotted in black with the green line representing the error spectrum, where the velocity zero-point corresponds to the optical depth-weighted median of {\MgII} ($z_{\rm abs}$). Regions of the spectrum that were not used to constrain the CMBM models are coloured grey.

Table~\ref{tab:EWs} lists the rest equivalent width of the strongest transition for a selection of the most commonly studied ions and the extreme velocity bounds of the absorption relative to $z_{\rm abs}$. In particular, the equivalent width of {\MgII}, $W_r(2796)=3.34$~{\AA}, classifies this system as an ultra-strong {\MgII} absorber \citep[$W_r(2796)\geq3$~{\AA}, e.g.][]{nestor07, nestor11, rubin10, gauthier13}. The bulk of the metal-line absorption is located within roughly $\pm225$~{\kms}, spanning a width of $\sim450$~{\kms}. Several weak clouds are more highly blueshifted at $\sim-400$~{\kms} and {\HI} also has absorption at $v\sim-650$~{\kms} although there are no associated metal lines with this region. It is clear that this system is multiphase, where the intermediate ions have broader kinematics than the low ions and the very blueshifted components are strongest in the intermediate ions, particularly {\SiIII}, {\CIII}, and {\CIV}.

The total column densities for a larger selection of ions are listed in Table~\ref{tab:columns} and were obtained from the metallicity modelling conducted in Section~\ref{sec:cmbm} (also see Appendix~\ref{app:metallicity}). The CMBM method described in the previous section differs from a traditional fitting method, e.g. where ratios of column densities from {\sc VPfit} are used to constrain ionisation conditions, metallicities, and densities. Instead, we are adjusting these physical model parameters as well as $N({\HI})$ and $b_{\rm nt}$ to produce acceptable fits to all of the line profiles. Only in this indirect way can we derive component column densities of the various ions. From this method, the DLA is best-modelled as having 30 VP components across two ionisation phases (low and intermediate). \citet{sameer2021, sameerleo} often fitted an additional collisionally-ionised phase constrained with {\OVI} or {\NV}. Both the {\OVI} and {\NV} doublets are covered in our spectrum but {\OVI} is heavily blended and {\NV} is not significantly detected. It is thus unlikely that a collisionally-ionised phase is dominant in this system. We did not model the components in {\HI} at $v\sim-650$~{\kms} because there are no associated metal lines, so their ionisation conditions are poorly constrained. The final VP component models are plotted in Figs.~\ref{fig:lowions} for the 20 components that are seen predominantly in low ionisation lines and \ref{fig:highions} for the 10 intermediate ionisation components. The total model to the data incorporating all 30 components and both the low and intermediate ionisation phases is shown as the orange line in Fig.~\ref{fig:absorption}.

\begin{table}
    \centering
    \caption{Total Rest Absorption Equivalent Widths and Velocity Bounds}
    \label{tab:EWs}
    \begin{threeparttable}
    \begin{tabular}{lccc}
    \hline
    Transition             & EW              & $v_-$\tnote{a} & $v_+$\tnote{a} \\
                           & ({\AA})         & ({\kms})       & ({\kms})       \\
    \hline
    {\HI}~$\lambda1026$    & $2.663\pm0.005$ & $-476$         & $+389$         \\
    {\HI}~$\lambda972$     & $2.144\pm0.004$ & $-476$         & $+261$         \\
    {\MgI}~$\lambda2853$   & $0.38\pm0.05$   & $-193$         & $+144$         \\
    {\MgII}~$\lambda2796$  & $3.34\pm0.01$   & $-474$         & $+223$         \\
    {\FeII}~$\lambda2600$  & $1.74\pm0.01$   & $-473$         & $+192$         \\
    {\SiII}~$\lambda1260$  & $1.557\pm0.003$ & $-473$         & $+222$         \\
    {\SiIII}~$\lambda1207$ & $1.553\pm0.005$ & $-460$         & $+232$         \\
    {\SiIV}~$\lambda1394$  & $0.63\pm0.02$   & $-454$         & $+196$         \\
    {\CIV}~$\lambda1548$   & $1.25\pm0.01$   & $-456$         & $+221$         \\
    \hline
    \end{tabular}
    \begin{tablenotes}
            \item[a] Absorption velocity bounds defined as the velocity at which the absorption recovers to within $1\%$ of the continuum. Only the extremes are reported, neglecting regions in between the extremes where the absorption also recovers to the continuum. The values are relative to $z_{\rm abs}=2.43014$.
    \end{tablenotes}
    \end{threeparttable}
\end{table}

For some transitions, discrepancies are present in the total VP models. The absorption in {\MgII}~$\lambda2803$ is underestimated, which appears to be a minor issue with the continuum fit since {\MgII}~$\lambda2796$ is generally well-modelled. There is also a slight overestimate in {\SiIV}, though comparing the column density with that from a model obtained with {\sc VPfit} suggests the difference is within uncertainties at $0.1$~dex. Since {\OVI} is not available due to blending and {\NV} is not detected, it is possible that some of the {\CIV} arises in an additional separate higher ionisation/warmer phase. If so then the {\SiIV} overproduction by the model would be reduced. Since the {\SiIV} is aligned with the {\CIV}, however, this contribution is not likely to be large. A slight abundance pattern difference from solar seems more likely, where the standard assumption of a solar abundance pattern is likely incorrect at some level for these clouds. The {\FeII} absorption is well-modelled, suggesting that the gas is not $\alpha$-enhanced. While it is not shown, there is a large underestimate in {\OI}, where the best-fit model provides a column density of $\log(N({\OI})/{\rm cm}^{-2})=15.34\pm0.03$ compared to a {\sc VPfit}-derived value of $17.05\pm0.45$. The abundance pattern may also be the cause of the {\OI} underestimate, or there are additional narrower components superimposed on the absorption, possibly indicating a more neutral phase that we do not currently account for. Since we are primarily interested in general trends in the absorption properties with the host galaxy properties, further investigation into abundance pattern variations and additional phases is beyond the scope of this work. 

Vertical lines in Fig.~\ref{fig:absorption} indicate the velocities of galaxies G1 (purple dotted), G2 (cyan dashed), G3 (green dot-dashed), and G4 (pink dot-dot-dashed) relative to the absorption. The shaded bars at the top of the panels represent the uncertainties on the galaxy redshifts from Table~\ref{tab:KCWIgalprops} and are stacked vertically to represent the relative distance of each galaxy from the quasar sightline, where G1 is the closest to the sightline while G3 is the furthest. The galaxies span the velocities covered by the primary kinematic region of low ionisation metal lines such as {\MgII}, {\FeII}, and {\SiII}, but not the full range covered by {\HI}, especially the {\HI} kinematic regions at $v\sim-400$~{\kms}, which have weak metal line components, and $v\sim-650$~{\kms}, where there are no obvious metal line components. The most massive galaxy, G1, is roughly centred on the optical depth-weighted median of the {\MgII} absorption ($z_{\rm abs}$) and G1, G2, and G4 have velocities roughly consistent with the kinematic region with the strongest absorption. G3 is furthest in velocity space from the strongest components, but closest to the weak components at $v\sim-400$~{\kms}. For the intermediate ionisation lines, {\SiIV} and {\CIV}, the galaxies roughly line up in velocity space with three primary kinematic regions (with an offset of $\sim50$~{\kms} for galaxies G1, G2, and G4). This suggests a possible connection between the metal-line kinematic regions and the individual galaxies, although kinematic alignment does not necessarily imply a physical connection, especially for galaxies further from the sightline.

\subsection {CGM Metallicity}
\label{sec:metallicity}

\begin{figure*}
    \centering
    \includegraphics[width=0.47\linewidth]{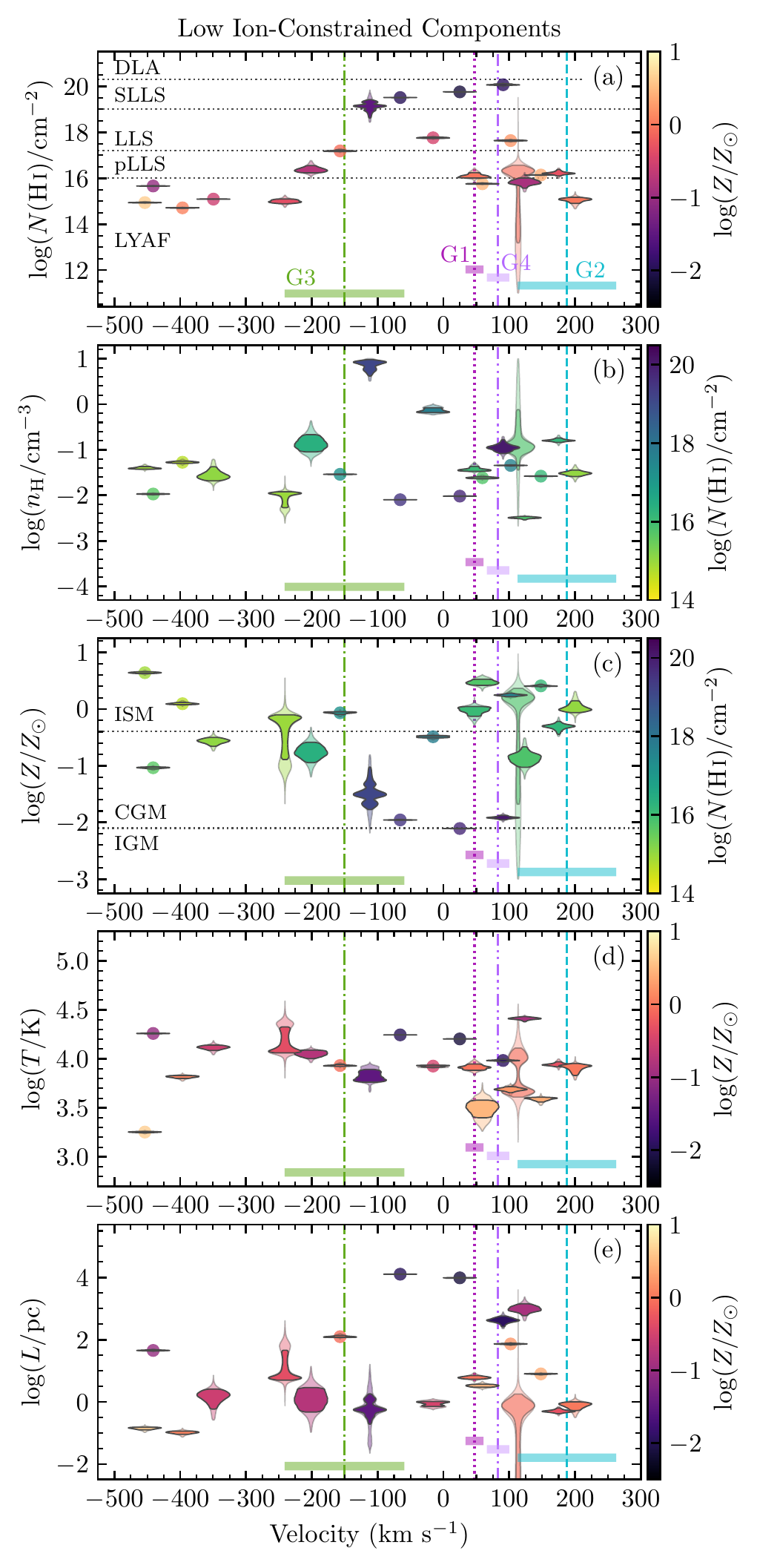}
    \includegraphics[width=0.47\linewidth]{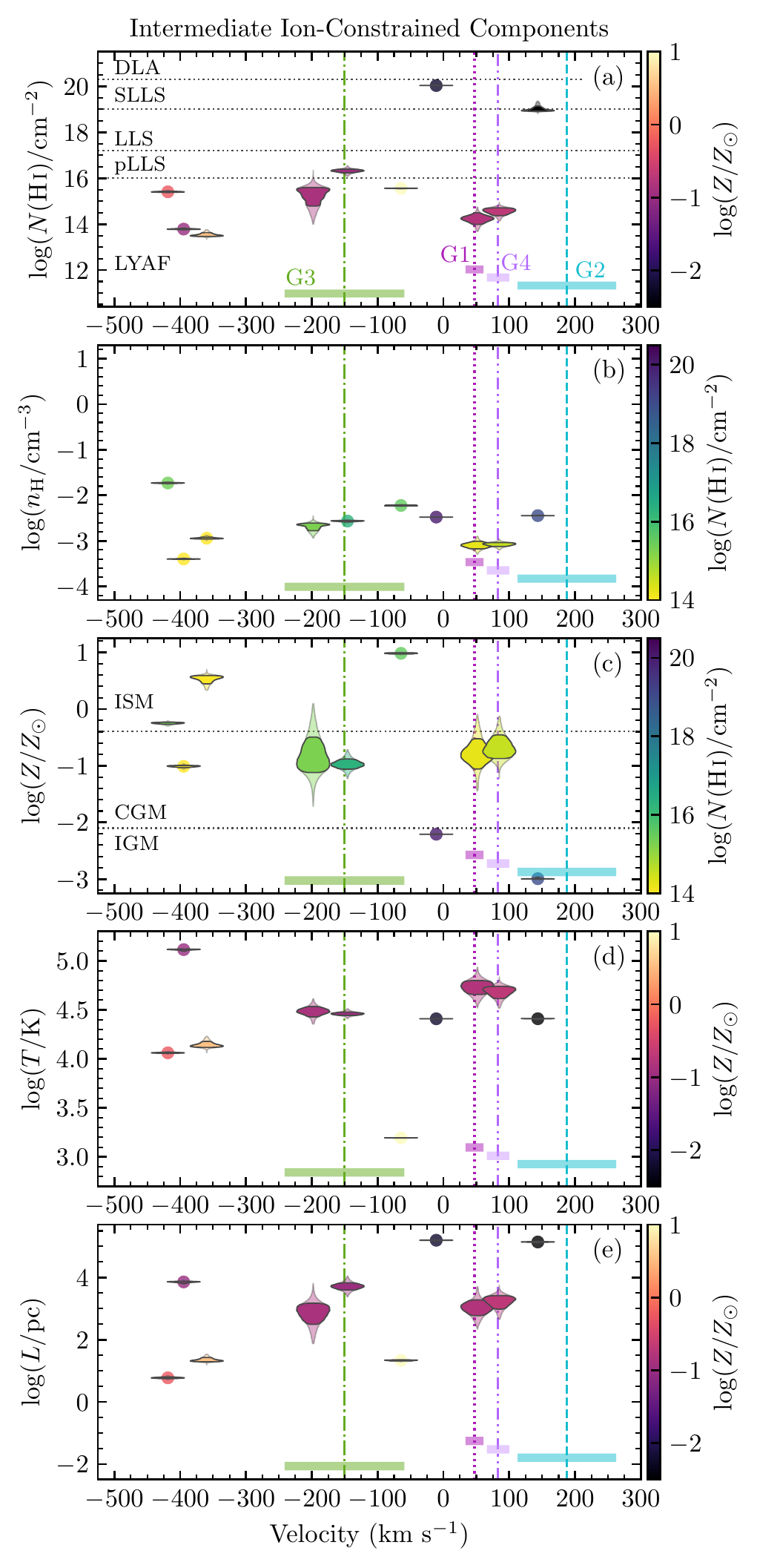}
    \caption{Violin plot showing the posterior distributions of (a) {\HI} column density, (b) hydrogen number density, (c) metallicity, (d) temperature, and (e) thickness for each ({\it left}) low ion-constrained component and ({\it right}) intermediate ion-constrained component as a function of velocity. Components are centred at their velocity relative to $z_{\rm abs}$. The full vertical range of each violin indicates the $3\sigma$ range of each y-axis value, while the darker part of the violin indicates the inner $1\sigma$ range. The widest part of each violin corresponds to the y-axis value with the highest probability for that component; the width of the violins is {\it not} representative of the component velocity uncertainties, which are not reflected here. Each violin is coloured by its (a, d, e) metallicity or (b, c) {\HI} column density. For the violins with small y-axis ranges, the colours are represented by circles at the median value. The highly uncertain low ion-constrained component is de-emphasised for clarity. Vertical lines and the shaded regions at the bottom of each panel represent the redshifts and uncertainties of galaxies G1 (purple dotted), G2 (cyan dashed), G3 (green dot-dashed), and G4 (pink dot-dot-dashed). The shaded regions are stacked to represent their relative distance from the quasar, where G1 is closest to the quasar while G3 is furthest away. Horizontal dotted lines in panel (a) delineate $N({\HI})$ for DLAs, super-Lyman limit systems (SLLS), LLS, partial LLS (pLLS), and {\Lya} forest (LYAF) systems \citep[e.g.][]{lehner16} while those in panel (c) represent rough metallicity boundaries for the ISM, CGM, and IGM.}
    \label{fig:violins}
\end{figure*}

The final component properties from the CMBM method including velocity, {\HI} column density, non-thermal $b$-parameter, hydrogen number density, metallicity, temperature, and cloud thickness are tabulated in Table~\ref{tab:cmbm}. The full posterior distributions for each parameter and component are presented in Fig.~\ref{fig:violins} as violin plots. In both figures, the full vertical range of each violin represents the posterior distribution ranges (with a $3\sigma$ sigma clipping) for each parameter on the y-axis, while the darker parts of the violins represent the $1\sigma$ distributions for comparison to typical literature values. The parameters for most components are well-constrained. The width of the violins represents the probability of that particular y-axis value, where the highest probability for each parameter and component is the y-axis value at which the violins are widest. Components are centred on their central velocity and coloured by either their {\HI} column density or metallicity. We separated the components constrained on the low ionisation lines from those constrained on the intermediate ionisation lines and also de-emphasised the low ionisation component with very large uncertainties for clarity. The horizontal dotted lines in panel (a) define the column density ranges for DLAs, super Lyman limit systems (SLLS), LLS, partial LLS (pLLS), and {\Lya} forest (LYAF) absorbers \citep[e.g.][]{lehner16}. In panel (c) they delineate the rough boundaries between ISM, CGM, and IGM metallicities. We used the $z\sim2-3$ mass-metallicity relation to estimate a boundary on the mean ISM metallicity of $12+\log({\rm O/H})=8.3$ for galaxies with $\log (M_{\ast}/M_{\odot})=8-10$ \citep[e.g.][]{erb06, steidel14, kacprzak15mmr}. The IGM boundary is the upper uncertainty of $\log(Z/Z_{\odot})=-2.85\pm0.75$ from the {\Lya} forest at $z\sim2.5$ \citep[e.g.][]{simcoe04}. We assume the values in between these boundaries are typical of CGM systems.

\begin{table}
    \centering
    \caption{Total Column Densities for the Best-Fit Model in Fig.~\ref{fig:absorption}}
    \label{tab:columns}
    \begin{tabular}{lcclc}
    \hline
    Ion      & $\log (N/{\rm cm}^{-2})$ & &  Ion     & $\log (N/{\rm cm}^{-2})$ \\
    \hline
    {\HI}    & $20.53_{-0.05}^{+0.03}$   & & {\OI}    & $15.34_{-0.02}^{+0.03}$ \\[3pt]
    {\MgI}   & $12.58_{-0.02}^{+0.02}$   & & {\NI}    & $14.48_{-0.02}^{+0.04}$ \\[3pt]
    {\MgII}  & $14.68_{-0.01}^{+0.01}$   & & {\NII}   & $15.32_{-0.01}^{+0.02}$ \\[3pt]
    {\SiII}  & $15.05_{-0.01}^{+0.01}$   & & {\NIII}  & $15.36_{-0.01}^{+0.01}$ \\[3pt]
    {\SiIII} & $14.75_{-0.01}^{+0.01}$   & & {\NV}    & $13.50_{-0.05}^{+0.04}$ \\[3pt]
    {\SiIV}  & $14.12_{-0.01}^{+0.01}$   & & {\MnII}  & $12.50_{-0.01}^{+0.02}$ \\[3pt]
    {\CII}   & $15.72_{-0.01}^{+0.01}$   & & {\FeII}  & $14.67_{-0.02}^{+0.02}$ \\[3pt]
    {\CIII}  & $15.95_{-0.01}^{+0.01}$   & & {\AlII}  & $13.90_{-0.01}^{+0.01}$ \\[3pt]
    {\CIV}   & $14.70_{-0.01}^{+0.01}$   & & {\AlIII} & $13.35_{-0.01}^{+0.01}$ \\[3pt]
    {\SI}    & $11.42_{-0.03}^{+0.04}$   & & {\PII}   & $13.02_{-0.01}^{+0.01}$ \\[3pt]
    {\SII}   & $14.54_{-0.01}^{+0.01}$   & &                                    \\
    \hline
    \end{tabular}
\end{table}

\begin{figure*}
    \centering
    \includegraphics[width=\linewidth]{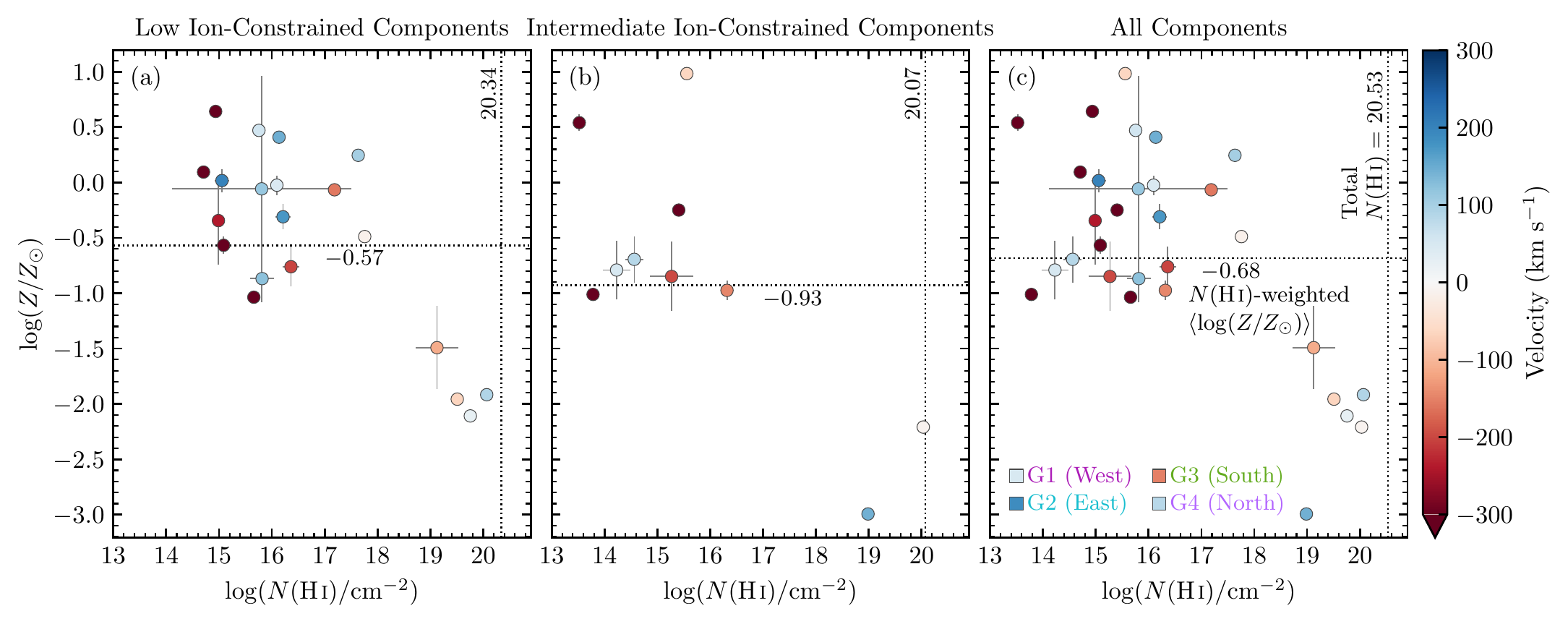}
    \caption{Component metallicity, $\log (Z/Z_{\odot})$, as a function of component {\HI} column density, $\log (N({\HI})/{\rm cm}^{-2})$, where the point colours correspond to the component velocity relative to $z_{\rm abs}=2.43014$ for (a) low ion-constrained components, (b) intermediate ion-constrained components, and (c) all components. The grey error bars represent the $1\sigma$ standard deviations, most of which are smaller than the points themselves. The dashed lines show the total {\HI} column density and $N({\HI})$-weighted mean metallicity for each set of components. There is a general trend of higher metallicity in the lower $N({\HI})$ components and lower metallicity in the higher $N({\HI})$ components. The low ionisation gas has a higher metallicity and higher {\HI} column density on average than the intermediate ionisation gas. For reference, the velocities relative to $z_{\rm abs}$ for galaxies G1 ($v=47$~{\kms}), G2 ($v=188$~{\kms}), G3 ($v=-150$~{\kms}), and G4 ($v=83$~{\kms}) are indicated by the coloured squares in the lower left.}
    \label{fig:ZNHI}
\end{figure*}

The components modelled to this DLA system span a wide range of properties. The column densities range from $13\lesssim\log (N({\HI})/{\rm cm}^{-2})\lesssim20$, corresponding to typical {\Lya} forest absorption strengths on the lower end up to nearly DLA strengths, although the total column density classifies this system as a DLA. The metallicities have a range of $-3\lesssim\log (Z/Z_{\odot})\lesssim1$, which suggests this gas is tracing both very metal-poor environments likely from accreting filaments and very metal-rich outflowing gas, with most clouds somewhere in between. Most clouds have thicknesses between $1-10,000$~pc and temperatures in the range $\log (T/{\rm K})=3.5-4.5$. Comparing the overall distributions of the components constrained on low ionisation lines to those constrained on the intermediate ionisation lines, we generally find higher hydrogen number densities, lower temperatures, and smaller thicknesses for the low ionisation components. Clouds in both ionisation phases span the full range of {\HI} column densities and metallicities and are found across the sightline in velocity space. This all suggests that there are several different physical structures along the line of sight and that the two phases are not necessarily physically coincident. There is still significant overlap in the properties of the components giving rise to the two phases, which perhaps indicates a spatial transition from the lowest ionisation gas towards higher phases.

We can also compare the component properties to the galaxies in the field. For both ionisation phases, a majority of the CGM clouds are consistent in velocity space with the galaxies in this compact group. The highest column density clouds are generally located within $\pm50$~{\kms} of a galaxy, with lower $N({\HI})$ clouds at higher velocity. Of the four low ionisation clouds with $\log (N({\HI})/{\rm cm}^{-2})>18$, one aligns with G1, one with G4, and two with G3 within the redshift uncertainties. The two strongest low ionisation clouds, which are associated with G1 and G4, are located within $v\sim10$~{\kms} of the galaxy redshifts. For the two intermediate ionisation components with strong $N({\HI})$, one aligns with G2 while the other does not have velocities consistent with any galaxies. In contrast, the lowest column density clouds for both ionisation phases are generally furthest from the galaxies in velocity space (being highly blueshifted in most cases). However, this may be due to either physical distance or simply peculiar motions from the gas flows around the galaxies. Focusing on the hydrogen number density panels, we find that the highest $n_{\rm H}$ components are associated with the low ionisation phase and are aligned with G1 or G3 kinematically. These components have small cloud sizes ($\sim 1$~pc) and low-to-average metallicities. Their density is the only outlying property compared to the rest of the clouds.

The lowest metallicity gas clouds primarily overlap with G1, G2, or G4 in velocity space for both phases, with one cloud aligning with only G3 due to its large redshift uncertainty. These clouds have metallicities $\log (Z/Z_{\odot})\sim-2$, and an intermediate ionisation cloud with a metallicity of $\log (Z/Z_{\odot})\sim-3$ is associated with G2. They are large, with thicknesses on the order of $10-100$~kpc, and have high column densities, but their densities and temperatures are average. These low metallicity clouds near the galaxies could be located in extended disks or embedded in tidal material being stripped from one or both of the two galaxies located closer to the quasar sightline. More likely given their very low metallicities, the clouds could also be accreting material falling on to this compact group as it is likely located at a node of multiple cosmic web filaments. It is unsurprising in both cases that we do not find extended disk material or accretion with velocities closely consistent with G3 since the gas would likely be disrupted by galaxies G1 and G2 on its way to G3 from the direction of the quasar sightline. 

In contrast, the highest metallicity low ionisation clouds are mostly clustered around galaxies G1, G2, and G4 or at high velocity away from all galaxies. The highly blueshifted clouds at $v\sim-400$~{\kms} are typically solar to super-solar metallicity and are found in both phases. These clouds are most likely associated with outflowing gas and may have originated from G1 since it is the most highly star-forming galaxy in the group, although G2 may also be capable of driving outflows. The highest metallicity cloud is an intermediate ionisation component that lies about halfway between the velocities of G1 and G3. Given that outflows can have maximum line-of-sight velocities up to $V_{\rm out}\simeq800$~{\kms} at this epoch \citep[e.g.][]{steidel10}, the alignment comparisons between the gas clouds and galaxies are not straightforward. Instead, connecting outflowing gas to host galaxies depends significantly on the SFRs and morphologies of the galaxies relative to the quasar sightline. Further interpretation of the gas is left to Section~\ref{sec:discussion}.

Figure~\ref{fig:ZNHI} presents component metallicities as a function of component {\HI} column densities for (a) low ion-constrained components, (b) intermediate ion-constrained components, and (c) all components. In general, we find a trend that higher column density components are more metal-poor, with $\log (Z/Z_{\odot})\sim-2$ for $\log (N({\HI})/{\rm cm}^{-2})>19$, and are tightly clustered in the $Z-N({\HI})$ plane. In contrast, the lower column density components are more metal-rich, with $-1.0<\log (Z/Z_{\odot})<1.0$ for $\log (N({\HI})/{\rm cm}^{-2})<18$, and are more loosely distributed. The total {\HI} column density for the system, calculated by summing the individual component contributions, is $\log (N({\HI})/{\rm cm}^{-2})=20.53$. Since most CGM studies report an ``average'' metallicity accounting for the total column densities in each ion, we also measured the $N({\HI})$-weighted mean metallicity for comparison (this literature value comparison is discussed in Section~\ref{sec:disc-metal}). This system has a sub-solar metallicity on average, with $\log (Z/Z_{\odot})=-0.68$, roughly consistent with the boundary between typical ISM and CGM metallicities. Comparing the components constrained on the low ions to those constrained on the intermediate ions, we find a higher total column density ($20.34$ and $20.07$, respectively) and more metal-rich gas ($-0.57$ and $-0.93$, respectively) for the low ions than the intermediate ions.

\section{Discussion}
\label{sec:discussion}

In the preceding section, we detailed the compact group environment associated with a DLA at $z_{\rm DLA}=2.431$, containing three galaxies with redshifts confirmed by their {\Lya} emission, G1, G2, and G3, and a fourth galaxy with no {\Lya} emission but confirmed via ISM absorption lines, G4. All of the galaxies are within $D=20-40$~kpc of the background quasar sightline, which also hosts rare ultra-strong {\MgII} absorption, and are within 30~kpc of each other. G1 and G2 may be capable of driving outflows based on their $\Sigma_{\rm SFR}>0.1$~M$_{\odot}$~yr$^{-1}$~kpc$^{-2}$ when assuming higher dust extinction values, whereas G3 and G4 have SFRs too low to drive outflows regardless of their dust extinction. G1 is the most massive galaxy in the group, having a luminosity of $1.66L_{\ast}$, and the other three galaxies are an order of magnitude fainter. While the morphology modelling of these galaxies is difficult due to their small size in the {\it HST} imaging and the possible presence of tidal features between galaxies, they mostly appear to have intermediate inclinations (G3 is the most face-on). G1, G2, and G3 and are largely probed along their major axis while G4 is probed most along the minor axis. 

The associated absorption is kinematically complex, spanning a primary velocity range of at least $450$~{\kms} with weaker absorption that is highly blueshifted at $v\sim-400$~{\kms} from G1. The absorption comprises 30 VP components (clouds) across low and intermediate ionisation phases, with cloud metallicities having a large range of $-3\lesssim\log (Z/Z_{\odot})\lesssim1$ and an average $N({\HI})$-weighted metallicity of $\log (Z/Z_{\odot})=-0.68$. A majority of the clouds align in velocity space with at least one of the four galaxies, but the highly blueshifted clouds do not. These results all point to a highly complex IGrM around this compact group, with multiple possible physical origins for each cloud of gas.

\subsection{Ultra-Strong Mg~{\sc ii}}

Focusing on {\MgII}, we find a large equivalent width of $W_r(2796)=3.34$~{\AA}, which classifies this system as an ultra-strong {\MgII} absorber \citep[e.g.][]{nestor07, nestor11, rubin10, gauthier13}. The strong equivalent width is a result of the large kinematic spread rather than a large column density. As discussed by \citet{nestor11}, the minimum rest-frame velocity width of a completely saturated absorber is 
\begin{equation} \label{eq:vmin}
\Delta v_{\rm min} = (\Delta \lambda / \lambda) \times c = (W_r(2796)/\text{\r{A}}) \times 107~{\kms},
\end{equation}
which gives $\Delta v_{\rm min}=357$~{\kms} for $W_r(2796)=3.34$~{\AA}. In contrast, the majority of the {\MgII} absorption for our system spans $\pm220$~{\kms} for a width of $\sim450$~{\kms} with additional weak absorption centred at roughly $-400$~{\kms} (total width $\sim700$~{\kms}). This system has a wider velocity spread by at least 100~{\kms} than expected for a purely saturated absorber (and even as much as twice as wide), suggesting that the absorption is made up of a complicated velocity structure with multiple components spread out across a large range of velocity rather than a single saturated component. The large number of components then suggests that each of the four compact group galaxies is giving rise to a separate portion of the observed CGM, the gas flows are quite dynamic in their motions, or some combination of the two. The compact group galaxies have redshifts spanning $\Delta v\sim340$~{\kms} and could be as large as $\Delta v\sim500$~{\kms}, which are comparable to the velocity spread of the bulk of the absorption. While the column density may be less important for determining the equivalent width of this system than the kinematics, column density depends on the total path length probed as well as the metallicity, ionisation conditions, and temperature of the gas, where these properties are undoubtedly important in determining the physical origins of the gas. 

For comparison, the total {\MgII} column density of this system, $\log (N({\MgII})/{\rm cm}^{-2})=14.68$, is similar to that of {\MgII} absorbers at $z<1$, where \citet{magiicat4} have a mean column density of $\log (N({\MgII})/{\rm cm}^{-2})=14.26$ ($1\sigma$ standard deviation of 1.1) for their sample of 47 isolated absorber--galaxy pairs with $W_r(2796)\lesssim2$~{\AA}. Ignoring any possible redshift evolution, the compact group environment likely does not have a strong influence on the {\MgII} column density and this is consistent with \citet{magiicat6} who found that {\MgII} column densities in group environments are consistent with those in isolated galaxies. In contrast, \citet{magiicat4} measured maximum velocity widths of $\sim200$~{\kms} for star-forming isolated galaxies at $z<1$ and $W_r(2796)\lesssim2$~{\AA}, which is roughly half that found here for the bulk of the absorption. A more consistent velocity comparison can be made with group environments, where the average {\MgII} profile in groups has a similar maximum velocity width as the isolated sample ($\sim200$~{\kms}) but the groups have more optical depth at these larger velocities than isolated galaxies \citep{magiicat6}. The result that our ultra-strong {\MgII} absorber is in a compact group environment may then be unsurprising. 

These comparisons above are for lower redshifts when galaxies do not drive the strong outflows that are ubiquitous at Cosmic Noon \citep[e.g.][]{steidel10, rupke-review}. At $z=2-3$, the {\MgII} absorbers in the \citet{chensimcoe17} sample have typical velocity widths of $\sim240$~{\kms} for all equivalent widths (slightly larger than at $z<1$) and $\sim580$~{\kms} for ultra-strong {\MgII} absorbers with $W_r(2796)\geq3.0$~{\AA}. The galaxies have not yet been identified for their sample, so it is unclear what types of environments their absorbers trace, but the authors suggest that the large velocity widths are tied to the heightened star formation rates at this epoch. \citet{bondwinds} also suggested that ``superwinds'' give rise to the strongest {\MgII} absorbers. It is possible that the large velocity width we see in this DLA system is primarily due to outflowing gas at the epoch of peak star formation. However, while G1 and G2 are star-forming and likely give rise to some metal-rich outflowing clouds in the absorber, none of the galaxies is currently star-forming enough to drive large amounts of outflowing gas (assuming typical dust extinctions). The large mix of low and high metallicity gas in the clouds along this sightline also do not point to outflows as being the dominant physical process giving rise to this gas, especially the very metal-poor $\log(Z/Z_{\odot})\lesssim-2$ clouds, which have the largest $N({\HI})$ values.

Given the compact group environment, the large {\MgII} kinematic spread could also be a result of galaxy--galaxy interactions. \citet{bordoloi11} suggested that a superposition of halos explained the elevated equivalent widths found in group environments at $z<1$, which they interpreted to mean that group environments did not influence the CGM. However, \citet{magiicat6} found that the velocity spreads in such a scenario were predicted to be much larger than those observed in their sample of loose groups due to the relative velocities between galaxies. The latter authors suggested that the CGM of the galaxies in group environments had already coalesced into an overall IGrM. The velocity spreads predicted for their superposition model and sample, $\Delta v<600$~{\kms}, are surprisingly comparable to the velocity spread of the ultra-strong {\MgII} absorption presented here, which could point to individual galaxy contributions to the absorption for our compact group. The majority of the \citeauthor{magiicat6} group environments are loose groups, with less than five galaxies in each group (most with only two galaxies) and typical separations between galaxies being $\gtrsim 60$~kpc. The compact group we study here has a maximum impact parameter of $\sim30$~kpc between the most distant group members, which is significantly smaller than the \citeauthor{magiicat6} sample. It is likely the more compact environment presented here results in stronger, more complicated absorption than found for more loose group environments in the form of both increased star formation-induced outflows and tidal stripping from more active galaxy--galaxy interactions. Additionally, the quasar sightline is much closer to all four compact group members ($D=20-40$~kpc) than the quasars in the loose group samples, such that the sightline here likely traces gas that is more likely to be associated with individual galaxies rather than an IGrM at larger distances.

Compared to other ultra-strong {\MgII} absorbers with spectroscopically-confirmed hosts, our system is comparable in velocity width. We found $\Delta v\sim450$~{\kms} for the bulk of the absorption and $\Delta v\sim700$~{\kms} including the highly blueshifted clouds. Other authors found values ranging $\Delta v=390-1000$~{\kms} at $z\sim0.5-0.8$. While most of the other works suggested that their absorbers originated from a starburst, three of the five systems were in loose groups \citep{nestor11, gauthier13}. Another system is a recent merger \citep{rubin10} that later had follow-up KCWI observations of the {\MgII} emission, where \citet{burchett21} suggested that the gas traced an isotropic outflow. This latter result is interesting because, while this merger is at $z\sim0.7$, observations and simulations both suggest that outflows at $z\sim2$ are isotropic \citep{steidel10, nelson19}, so similar processes may be at play in this system as in ours. The fifth ultra-strong {\MgII} absorber in the literature, while it is at $z\sim2$ unlike the other lower redshift work, is associated with a very low impact parameter ($D=0.9$~kpc) galaxy \citep{noterdaeme12}. This is a different type of system than the one through our compact group since the quasar sightline is likely piercing the disk of the galaxy itself, although the authors did not survey that field beyond 10~kpc so it is unclear what type of larger environment that galaxy resides in. These previous works all considered the absorption as a single physical structure, which, from our photoionisation modelling and the large velocity spreads discussed here, is too simple for determining the physical origin of the gas. We continue to explore this in the next sections.

\subsection{Metallicity}
\label{sec:disc-metal}

Our photoionisation modelling detailed in Section~\ref{sec:metallicity} revealed a complex distribution of cloud metallicities along the sightline, which is not reflected in the sightline-averaged metallicity that is typically reported in the literature. Previous authors fitted a single DLA component with $\log (N({\HI})/{\rm cm}^{-2})\sim20.35$ and found average metallicities for this system ranging from, e.g., ${\rm [Si/H]}=-0.7$ to $-0.8$ \citep{lu98, noterdaeme07, wang2015}, ${\rm [S/H]}=-0.6$ to $-0.9$ \citep{dodorico2002, d-z2004, noterdaeme07, lehner14}, and ${\rm [Zn/H]}=-0.9$ \citep{ledoux06, noterdaeme07}. In comparison, we found a total $\log (N({\HI})/{\rm cm}^{-2})=20.53$ and $N({\HI})$-weighted average metallicity of $\log (Z/Z_{\odot})=-0.68$, which is comparable to the ranges measured previously. The CMBM method for characterising the metallicity of the gas along the line-of-sight then recovers the average metallicity found in previous works in addition to providing a more detailed view of the gas properties \citep[this is true in simulations as well;][]{marra22}. \citet{sameer2021, sameerleo} also recovered the average metallicities for lower {\HI} column density systems, which, when combined with the analysis here, suggests that the CMBM method works for a wide range of absorbers. 

While we found an $N({\HI})$-weighted median metallicity consistent with previous works, we note that these works did not all account for all of the gas and ionisation phases present in the system. For example, \citet{d-z2004} fitted a single DLA component to the {\HI} in the VLT/UVES spectrum, which neglects over half of the velocity spread of the {\HI} absorption in the system, where the saturated trough of their DLA component spans roughly the redshift range of galaxies G1, G2, and G4 ($v\sim0-200$~{\kms}) and neglects the absorption at negative velocities relative to $z_{\rm abs}$. It appears that they then associated all of the metal lines to this single DLA component. Their metallicity value thus fully accounted for all of the metal line absorption in the system, but less than half of the velocity coverage of {\HI}. In some cases, previous works only modelled the low ionisation phase while higher phases traced by ions such as {\SiIV} and {\CIV} are neglected \citep[e.g.][]{noterdaeme07} even though our analysis demonstrates that a large fraction of the {\HI} is associated with this phase. Our analysis highlights the need to account for all of the absorption for gas interpretations.

More broadly, our average metallicity for this absorption system is on the upper side of typical DLAs at $z=2-3$ \citep{rafelski12} and is typical of DLAs with spectroscopically-confirmed host galaxies at $z\sim2$ \citep[for an overview, see][]{krogager17}. However, the impact parameter of our DLA-host group is larger than the \citeauthor{krogager17} compilation, with $D\sim30$~kpc ($D=21.5$~kpc for our nearest galaxy) compared to their mean $D=8.3$~kpc. The majority of the galaxy searches were done for metal-rich DLAs because it was assumed that DLAs follow a mass--metallicity relation, where more metal-rich DLAs should be hosted by massive/luminous galaxies that are more easily observed at these redshifts. Many of the surveys also focus on a small region ($\sim10$~kpc) around and on top of the quasar sightline since absorber column densities are generally anti-correlated with impact parameter. The current samples are then biased to very low impact parameters and higher mass galaxies. More recently, \citet{mackenzie19} used MUSE to identify host galaxies of several $z\sim3.5$ DLAs with a wider range of metallicities and field-of-view. They found several galaxies (with varying confidence) in five out of six fields at the DLA redshifts, with impact parameters $25-280$~kpc, where their most metal-poor DLA ($\log(Z/Z_{\odot})=-2.33$) was associated with three galaxies in a large-scale filament-like structure. The authors suggested that at most only two of their identified galaxies in their full sample are the likely DLA hosts, with the rest lying below their detection sensitivity at lower impact parameters. 

The DLA-hosting compact group of galaxies we identified here is thus an outlier in impact parameter, number of galaxies, and the group compactness relative to the current $z=2-3$ DLA host galaxy literature. Our system may also be an outlier on the DLA line width--metallicity relationship \citep[e.g.][]{neeleman13}, having a lower metallicity than expected from the full {\SiII} line width ($\sim700$~{\kms}) at this redshift, though it is roughly at the lower envelope of the relation if we neglect the highly blueshifted weak clouds (line width $\sim450$~{\kms}). The system is also on the lower envelope of the {\MgII} equivalent width--metallicity relationship \citep[][however, considering the velocity spread, this system would still be an outlier]{murphy07}. It is possible that the lower average metallicity relative to the metal line widths and strengths for this DLA could be due to significant amounts of metal-poor accretion where the compact group sits at a cosmic web node. This would be consistent with the large-scale structure found at the DLA redshifts in the \citet{mackenzie19} survey and with another line width--metallicity outlier studied by \citet{fu21}, although the latter host galaxy is a massive starburst, unlike our compact group. It is difficult to say for certain if the lower metallicity DLAs that have thus far been mostly neglected in host galaxy searches should be found in more dense environments since the current samples are too small. We further explore the possible sources of the gas in the next section.

While we compared the average metallicity for our compact group DLA to the literature, it is clear that our more detailed metallicity analysis provides insight into the many structures present in the CGM. Specifically, we found that the clouds along the line-of-sight have metallicities ranging over nearly 4~dex ($-3 \lesssim \log(Z/Z_{\odot}) \lesssim 1$) across roughly 700~{\kms}. The average metallicity for this sightline then neglects the diversity of structures being probed by the background quasar and muddies the interpretations for the physical origins of the gas. This metallicity range is larger than that found in \citet{crighton15} using a similar analysis, although their system is in a different environment than ours, where the absorption was associated with a presumably isolated galaxy. The authors interpreted their $z=2.5$ Lyman limit absorber as being an outflow from the galaxy since the clouds are more metal-rich with a range of $-1.0<\log(Z/Z_{\odot})<-0.2$ across roughly 440~{\kms}. A similarly narrow range of metal-rich clouds (at most 1~dex, $-0.8\lesssim\log(Z/Z_{\odot})\lesssim0.2$) was found for several kinematic regions across over 300~{\kms} of a $z=2.32$ DLA hosted by another presumably isolated galaxy \citep{bouche13}. Because the total metallicity of their DLA was significantly below the host galaxy's ISM metallicity, the authors assumed the gas was a combination of accretion and rotating disk components due to the major axis sightline. Our larger metallicity range in comparison could be a result of the galaxy environment or our method of also accounting for the multiphase structure of the gas, which the other works did not consider. 

Simulations have demonstrated that while there may be, for example, both low metallicity accretion and high metallicity outflows along a given line-of-sight, they often line up in velocity space, and the complexity of the gas in the CGM is more clear with increasing resolution \citep[e.g.][]{churchill15, hafen19, hummels19, peeples19, vandevoort19, marra22}. In FOGGIE, the simulated sightlines pass through a wide range of gaseous environments within a single halo, with absorbing gas clouds tracing metallicities with rough ranges comparable to our clouds \citep{peeples19}. Specifically, one of the simulated sightlines they presented hosts a $\log (N({\HI})/{\rm cm}^{-2})=16.82$ system where the gas along the line-of-sight ranges in metallicity from $\log(Z/Z_{\odot})=-3$ to super-solar values at $z=2.5$ and similar variations are present in other simulated sightlines. While the authors selected their halo to have a quiescent merger history at $z<1$, it appears to have multiple galaxies at $z=2$, which could be giving rise to the rich gaseous environment in the CGM on some level. We expect that such a large range of cloud metallicities might be a common feature of compact group sightlines, especially at low impact parameters.

\subsection{Possible Gas Origins}

Our detailed metallicity modelling using CMBM has afforded us the ability to detect clouds along the line-of-sight with very different properties that are often washed out when the absorption system is treated as a single structure. Given the large variation in gas properties along the line-of-sight through our compact group, there are several possible physical origins for the clouds, which could all be present at the same time. Here we describe these possibilities, including star formation-driven outflows, accretion onto the group environment from the cosmic web, tidal streams from interactions between the group member galaxies, and dwarf satellite galaxies below our detection limit.

\subsubsection{Metal-Rich, High Velocity Outflows} 

Outflows are expected to be ubiquitous at $z=2-3$ since the SFR density peaks at these redshifts \citep{madau14}. Indeed, observations at this epoch regularly observe strong outflows with velocities exceeding several hundred kilometres per second \citep[e.g.][]{steidel10, rupke-review} and they may be isotropic around galaxies rather than having bipolar morphologies since the SFRs are higher and disks are not as well-formed as at lower redshifts \citep[e.g.][]{law12, nelson19}. For our system, the clouds that are most highly likely to be outflowing material are those that are blueshifted with velocities $v\sim-400$~{\kms}. From Fig.~\ref{fig:violins} we see that both the low and intermediate ion clouds at these velocities have low column densities, metallicities greater than ten percent solar (most are consistent with ISM metallicities), and do not overlap in velocity space with any of the four galaxies. The combination of low column density and high velocity is similar to the comparable components fitted to {\MgII} at $z<1$ in \citet{magiicat5}. The authors found that their high velocity, low column density components were associated with face-on galaxies and minor axis sightlines, which are orientations where outflow signatures should dominate over other gas flows. While G3 is the most face-on galaxy in the group and nearest to these high velocity clouds in velocity space, it is the furthest galaxy from the quasar sightline and has the lowest SFR, so it is unlikely to be giving rise to this gas. G4 has the most minor axis sightline of the group, but again, it is not highly star-forming enough to eject these clouds at such high velocities. Both G1 and G2 make good candidates for outflowing gas sources since they are both star-forming (although not enough to drive significant outflows assuming average dust extinction; $\Sigma_{\rm SFR}>0.1$~M$_{\odot}$~yr$^{-1}$~kpc$^{-2}$) and are close to the quasar sightline where the other group members are less likely to disrupt the outflowing gas before it is intersected by the quasar.

Since G1 is the galaxy closest to the quasar sightline and has the highest SFR in the group (it is on the star-forming main sequence at this epoch), it is most likely that the highly blueshifted clouds are outflowing gas from this galaxy. This would mean that these clouds have a minimum outflow velocity of $v_{\rm out}>400$~{\kms} from G1. Our morphology modelling indicates that G1 is highly inclined with $i=66^{\circ}$, so the outflow velocity could be as large as $v_{\rm out}\sim1000$~{\kms} after correcting for inclination. This value is comparable to the $v_{\rm out}\simeq800$~{\kms} outflows estimated from stacked galaxy spectra at $z=2.3$ \citep{steidel10} and is the upper bound placed on the outflow inferred from modelling the {\MgII} absorption hosted by another star-forming main sequence galaxy in another field in our $z\sim2$ KCWI survey \citep{nielsen20}. Since we do not have masses for the galaxies in our compact group, we cannot say for certain if this outflowing gas will remain bound within the group or escape its potential well, but the material is likely to accrete onto the other galaxies at a later epoch given their close proximity.

Focusing on the metallicity, we find that roughly half of the highly blueshifted clouds have solar to super-solar values, which is consistent with typical ISM metallicities at this epoch (estimated from the mass--metallicity relation). These clouds may not yet have encountered a significant amount of lower metallicity CGM material as they have moved outward from the galaxy since they are quite high metallicity (above the ISM dotted line in Fig.~\ref{fig:violins}) and not diluted. The rest of the clouds have metallicities above ten percent solar, which suggests that they may be entrained material that was already within the CGM (below the ISM line in Fig.~\ref{fig:violins}) and is being pulled along by the outflow. Depending on the galaxy masses, which we are not able to estimate here, these clouds are slightly more metal-rich than the outflowing gas predicted in the FIRE simulations by \citet{hafen19} but are comparable to the outflowing metallicities predicted by e.g. \citet{vandevoort+schaye12}, \citet{shen13}, and \citet{nelson19} at $z=2-3$. Compared to observational results, these values are much larger than the ${\rm [Si/H]}=-1.5$ estimated by \citet{nielsen20} for a minor axis sightline $\sim70$~kpc from an edge-on galaxy at $z\sim2$, which they inferred was an outflow that had entrained significant amounts of CGM gas. However these latter authors reported a total metallicity that washes out the line-of-sight structure. Using similar photoionisation modelling methods, \citet{crighton15} found similar cloud metallicities as ours, which they also inferred originated from an outflow. These results strengthen the interpretation that these highly blueshifted clouds are an outflow, likely from G1 either recently (if G1 is more dusty) or from a star formation episode that has since slowed (if G1 has a typical dust extinction).

It is interesting that there is no comparable set of highly redshifted clouds in this system, especially if outflows at this epoch are generally isotropic. If we assume that only G1 is driving high velocity outflows, then this could be explained by the outflow having a small opening angle (i.e. not isotropic) and the combination of G1's large inclination and major axis azimuthal angle, all of which would make it so that the quasar sightline can only probe blueshifted gas. However, G2 may also be capable of driving outflows and it has similar inclination and azimuthal angles as G1. G2 could be driving lower velocity outflows where the clouds overlap in velocity with the other galaxy redshifts, such as the clouds at $v\sim-200$~{\kms} and the very high metallicity intermediate ion-constrained cloud at $v\sim-65$~{\kms}. These clouds would have outflow velocities ranging from $v_{\rm out}=120-300$~{\kms} at a minimum and $v_{\rm out}=230-570$~{\kms} if corrected for G2's inclination ($i=58^{\circ}$). Since G2 is less massive than G1, it would likely have a lower ISM metallicity from the mass--metallicity relationship so any outflowing clouds could also be less metal-rich than the highly blueshifted clouds. The slightly larger impact parameter and lower SFR of G2 could also mean that any outflows it is ejecting will have less of a chance of reaching the quasar sightline. Because of these factors, it is more difficult to clearly identify blueshifted outflowing clouds originating from G2, but they may still be present in Fig.~\ref{fig:violins}.

\subsubsection{Metal-Poor IGM Accretion}

The high outflow rates found at this epoch and the fact that all four galaxies are star-forming (two out of the four compact group galaxies may even be capable of driving outflows) imply that the galaxies must also be accreting significant amounts of gas from the IGM to fuel the active star formation. Indeed, simulations predict that the accretion rate of gas onto galaxies is at its highest at this epoch \citep{vandevoort11} and that cold mode accreting streams may be best observed by searching for high {\HI} column density absorbers at low impact parameters to galaxies \citep{vandevoort+schaye12}, although the covering fraction of such gas is still likely small \citep[$3-4$ percent for DLAs;][]{fg11}. \citet{wright21} examined accretion in the EAGLE simulations where they traced particles classified as being ``first-infall'' (never processed through a galaxy) and ``pre-processed'' (previously in a galaxy), finding metallicities of $\log(Z/Z_{\odot})\leq-2$ and $\sim-1$, respectively. Observationally, accretion is often invoked for high {\HI} column density, low metallicity absorbers \citep[e.g.][]{fumagalli11, fumagalli16accretion, rauch11, ribaudo11, bouche13, borthakur19, zahedy19accretion, fu21}. A direct detection still remains difficult, but observing the CGM in emission is also a promising avenue for finding accretion \citep[e.g.][]{dcmartin19, cameron21}. 

For our compact group system, we find several high column density, low metallicity clouds along the line-of-sight in both ionisation phases and these components are aligned in velocity space with the galaxies nearest to the quasar sightline (G1, G2, and G4). We suggest that these components, which have $\log (N({\HI})/{\rm cm}^{-2})\gtrsim19$ and metallicities $\log (Z/Z_{\odot})\lesssim-2$ (including one cloud at $\log (Z/Z_{\odot})\sim-3$ in the intermediate ions), are tracing cold mode accreting filaments from the IGM. \citet{simcoe04} estimated IGM metallicities from the {\Lya} forest at $z\sim2.5$ and found metallicities of ${\rm [M/H]}=-2.85\pm0.75$, where the upper bound on this value is plotted as the horizontal dotted line delineating the CGM and IGM in Fig.~\ref{fig:violins}(c). If our observed low metallicity clouds are IGM accretion, then they are comparable to or slightly more metal-rich than the {\Lya} forest absorbers and the values found in simulations. \citet{vandevoort+schaye12} suggested that cold mode accretion could have fairly high metallicities near galaxies due to the gas mixing with outflowing material, or could simply be an artefact of their simulation resolution, where the metallicity had not yet converged. Regardless, it is less likely that these components are tidal streams or recycled accretion from outflows originating in the neighbouring galaxies because their metallicities are several dex lower than those expected for the ISM (upper dotted line in the figure) and still lower than typical CGM or ``pre-processed'' clouds at this epoch. 

The low velocity of these clouds, with several being located within $v\sim20-200$~{\kms} of one of the four galaxies' redshifts, may seem counter to predictions that accreting gas should align with galaxy discs to add angular momentum and therefore have relative velocities on the order of galaxy rotation \citep[e.g.][]{steidel02, ggk-sims, stewart11, stewart17, danovich12, danovich15, ho17, zabl19}. However, the compact group environment probably disrupts the gas as it accretes onto the wider IGrM, where the galaxies themselves are rapidly changing their locations within the group due to interactions. Furthermore, the group halo likely has a larger virial temperature than would be found around these galaxies individually, so it is unclear if the presumably accreting gas actually reaches the galaxies themselves as cold mode accretion. At best, the accreting material is probably reaching the overall group halo at $20-40$~kpc (minimum projected distance) but its fate within that distance is unknown. This gas must eventually reach the galaxies as some form of accretion since they are currently star-forming.

\subsubsection{Tidal Streams from Galaxy Interactions}

Given the close proximity of the four galaxies to each other and the weak evidence of faint, narrow, elongated streams of material coming off G1 and G4 in the {\it HST} imaging, it is highly likely these galaxies are actively interacting and being tidally stripped. In fact, simulations suggest that 96 percent of compact groups at $z=2$ are fully merged into a single massive galaxy by $z=0$ \citep[e.g.][]{wiens19}, so we may be observing the IGrM in the early stages of that merger process. At low redshift, \citet{chen1127} used VLT/MUSE to observe the CGM in emission for a group of 14 galaxies associated with a known DLA in a background quasar spectrum. Their emission mapping found a large intragroup nebula tracing the galaxies and their motions in the group, where they attributed a dense filament of gas nearest to the quasar sightline to ram-pressure and tidal stripping from galaxy--galaxy interactions. Their interpretation is consistent with previous work examining only the quasar absorption \citep{ggk1127, peroux1127} and is similar to other DLA hosts, where interactions between galaxies appear to give rise to the DLAs \citep[e.g.][]{battisti12, borthakur19}. In the local Universe where {\HI} can be directly imaged in great detail, \citet{deblok18} created a deep map of the {\HI} ($3\sigma$ column density limit $\log(N({\HI})/{\rm cm}^{-2})=19.1$) around the interacting M81 galaxy triplet, including starbursting galaxy M82. Nearly all of the detected {\HI} external to the galaxies in this group was associated with tidal streams. The distances between the galaxies in the M81 triplet are quite a bit larger than those found in our compact group, but the interactions between galaxies are expected to be similar.

Since ISM metallicities at $z\sim2$ can be as low as ${\log (Z/Z_{\odot})\sim-0.7}$ and it is ISM material that would be pulled out of galaxies into streams from galaxy--galaxy interactions, we suggest that many of the clouds in Fig.~\ref{fig:violins} are tidal stream material. In particular, the $\log(Z/Z_{\odot})>-1$ clouds with low velocities relative to galaxies G1, G2, and G4 seem most likely, where tidal material should retain the rough redshift of the host galaxy it is being stripped off. Ideally, we would have measured the rotation curves of each of the galaxies in this group to gain further insight into which galaxy each of these clouds might be most closely related since their relative velocities should continue to align with the rotation curve of the host galaxy. However, the KCWI spaxels have sizes on the order of galaxy sizes at $z\sim2$ \citep{allen17}, so we cannot measure the rotation curves for comparison. Regardless, galaxies at this epoch have rotation velocities on the order of up to a couple hundred kilometres per second, depending on their mass, and velocity dispersions of $\sigma\sim50$~{\kms} \citep[e.g.][]{wisnioski15}. These rotation velocities are larger than the offsets between the absorption clouds in question and galaxies G1, G2, and G4. However, we have not corrected for inclination effects in Fig.~\ref{fig:violins}, where these galaxies are highly inclined, so the corrected absorbing cloud--galaxy offset velocities could be larger and more consistent with the rotation velocities expected. Additionally, the group environment likely interrupts clear signatures of rotation out to large distances from the group member galaxies, especially for gas that was stripped at earlier times. It may be more likely that these stream clouds trace the wider group potential well and will eventually become part of the wider IGrM. In this last case, the tidal streams should not exceed the group dynamics defined roughly by the galaxy redshifts. Consistent with this, the bulk of the absorption, and in particular the clouds that are most likely associated with tidal streams, reside within the velocity space defined by the galaxy redshifts. In contrast, the clouds that are most confidently outflows reside outside the galaxies' velocity space.

Other potential clouds that could be tidal material are the two clouds at $v\sim-10$~{\kms} and $-110$~{\kms} with large hydrogen number densities that are outliers in the sample. While these clouds are not particularly metal-rich (the latter cloud is quite metal-poor, in fact) their densities suggest that this material was recently located in more dense environments than the CGM. It is possible that their lower metallicities mean that either these clouds have swept up surrounding metal-poor CGM material but have retained their dense structures or these clouds were stripped at an earlier epoch than the rest of the clouds, when the galaxy ISM metallicities would have been more metal-poor. Alternatively, their very high hydrogen number densities and parsec-sized thicknesses could also suggest that these clouds trace what could eventually be star formation regions in tidal streams, similar to those found in the nearby universe \citep[e.g.][]{ryanweber04, werk08, howk18a, howk18b}.

\subsubsection{Dwarf Satellite Galaxies}

Surveys searching for DLA host galaxies at high redshift often result in non-detections in {\Halpha} and {\Lya} \citep[for a summary, see][and references therein]{krogager17}, where faint dwarf galaxies below the survey detection limits are commonly invoked. While we have identified four host galaxies located in a compact group for our $z_{\rm DLA}=2.431$ DLA, it is still possible that dwarf satellite galaxies below our KCWI detection limits are present in the group. An earlier survey of the field around our DLA that did not cover our compact group resulted in no {\Halpha} emission detections from host galaxies within $D\sim12.5$~kpc and with ${\rm SFR}>4.4-11$~M$_{\odot}$~yr$^{-1}$, where the higher SFR limit is located within $0\farcs5$ ($\sim4$~kpc) of the quasar sightline \citep{wang2015}. In comparison, the galaxies we identified in the compact group have ${\rm SFR}_{\rm FUV}=1.4-11.6$~M$_{\odot}$~yr$^{-1}$, where most are lower than the detection limit in the previous authors' survey. From the mass--metallicity relation, such faint dwarf galaxies would have ISM metallicities that are lower than the more massive galaxies we observed here so it is unlikely they would be giving rise to most of the very metal-rich clouds. Moreover, dwarf galaxies in a compact group would be more likely to be impacted by ram pressure stripping as they move through the wider IGrM and around the more massive galaxies, so while they could be the source of gas in a gas stripping scenario, it is less likely that they are able to retain enough of their ISM for significant star formation activity to drive outflows \citep[for a review, see][]{cortese21}. Possible clouds that could be dwarf galaxy candidates are the very high hydrogen number density clouds discussed in the previous section, where the galaxies have retained enough gas to be detected in the quasar sightline but the galaxies themselves could be below our surface brightness limits. The low metallicity of these clouds could then be explained by the low expected mass of such objects, especially if they are tidal dwarf galaxies forming out of tidally-stripped or accreting material \citep[e.g.][]{leewaddell12}. However, the small sizes of these clouds, $\sim1$~pc, suggest this may not be very likely since this is an order of magnitude smaller than the tidal dwarf galaxy discovered by \citet{leewaddell12}.

\subsection{What does this mean for Compact Groups?}

At low redshift, compact groups often show evidence of star formation suppression, with a higher fraction of quiescent galaxies than in the field, and the individual galaxies themselves are often {\HI}-deficient, likely due to the continuous strong interactions between group member galaxies. \citet{verdes-montenegro01} suggested an evolutionary pathway using 72 of these environments where younger compact groups have more star-forming spiral galaxies that have retained their {\HI} reservoirs. Over time, these galaxies transition to more quiescent ellipticals that have had their {\HI} reservoirs stripped from repeated interactions and this stripped material is expected to reside within the surrounding IGrM. Consistent with this stripping scenario, \citet{borthakur10, borthakur15} found a diffuse component of {\HI} in the IGrM and this excess gas is found in increasing amounts with the evolutionary phases above. \citet{bitsakis16} used a much larger sample (1770 compact groups) to additionally suggest that the motions of the galaxies through the increasingly massive IGrM cause the ISM of the galaxies to shock and become more turbulent, both of which further suppress star formation. To fully test this evolutionary pathway, a multiphase view of the IGrM in addition to current observations of the galaxies themselves is needed. 

As we have shown here, quasar absorption line spectroscopy within $\sim30$~kpc of a compact group revealed a rich IGrM traced by a DLA and ultra-strong {\MgII} absorption, which are quite rare systems. The metallicities of the clouds along the line-of-sight clearly show that a variety of physical processes are active in this compact group, including outflows and tidal stripping, and suggest that the group itself is still actively accreting fresh IGM material, perhaps in contradiction to low redshift findings of {\HI}-deficient galaxies. While this is most likely a result of the higher redshift than the compact groups previously studied, this is the first time a compact group has had its IGrM studied in such great detail using quasar absorption lines,\footnote{Although Q1127$-$145 $z=0.313$  \citep{ggk1127, peroux1127, chen1127} and Q2126$-$158 $z=0.663$ \citep{whiting06} may classify as compact groups and have strong {\MgII}, their metallicities have not been studied in as much detail, if at all.} which are more sensitive to the diffuse intragroup gas than previous {\HI} surveys but probe a much smaller spatial region. From a sample of only one, it is unclear if all compact groups should show such rare complexity in their absorbing gas, where this more diffuse and more highly ionised absorbing material is otherwise ``invisible'' to {\HI} emission mapping, or if this is a feature of the redshift, evolutionary stage, or specific geometry of this system. Regardless, it is clear that there is a significant amount of diffuse gas in the IGrM around our compact group and this material could be shocking and disturbing the ISM of the lower luminosity (mass proxy) galaxy G3 in particular, suppressing its SFR. Obtaining a census of this material will clarify the role of galaxy--IGrM interactions in star formation suppression and determine whether the tidal material is easily ionised to higher phases, which is important for understanding the fate of the stripped gas \citep[for a discussion on this latter point, see][]{borthakur10, borthakur15}.

To further understand the evolutionary processes occurring in these environments, it would be best to compare the absorbing gas to {\HI} emission maps and estimates of {\HI} deficiencies in the galaxies for the various phases proposed by \citet{verdes-montenegro01}. Comparing the absorption ionisation phases and detailed metallicities to the galaxies and their {\HI} content for a sample of compact groups would provide a more complete picture of the physical processes at play and what factors are most important in giving rise to the observed low redshift {\HI}. While we do not have the ability to obtain {\HI} maps for this epoch to determine the evolutionary phase of this group, it is interesting that only 50 percent of the galaxies in this group may be star-forming enough to drive outflows (G1 and G2), but we still infer that the gas originates from outflows and accretion in addition to the tidal streams commonly observed at low redshift with {\HI}. This group may be in an earlier phase of the evolutionary pathway, especially if the metal-poor accreting gas is reaching these galaxies to continue fuelling their star formation, but the strong interactions between galaxies may be suppressing the SFR of G3 and G4 in particular. 

Connecting the absorption to {\HI} emission maps would also allow us to more confidently determine if the quasar is piercing the tidal streams directly \citep[e.g. Q1127$-$145 $z=0.313$ with {\Hbeta} emission;][]{chen1127} and, if so, whether that material is primarily being stripped from G3 and G4 to match their lower SFRs or is in the process of depleting G1 and G2 of their star-forming reservoirs. It is perhaps interesting that we do not observe a large {\HI} {\Lya}-emitting halo surrounding this group (see Fig.~\ref{fig:galspec}) since this would be an additional confirmation of a significant amount of {\HI} surrounding the group and even isolated galaxies at this epoch have been observed to host such structures \citep[e.g.][]{leclercq17, nielsen20, ychen21}. However, the physical mechanism giving rise to {\Lya} halos is still difficult to determine. One scenario might be that the lower SFRs of these galaxies mean that the photons required to illuminate the surrounding {\HI} are not being emitted by the galaxies rather than indicating a lack of {\HI} enclosing the group. Alternatively, our observations may not be deep enough to detect a {\Lya} halo. As stated above, we cannot conclude general characteristics from only one compact group, so a larger sample is required to further explore these results, but the large variation of absorption properties and the large kinematic spreads found for our system may be more common in these sorts of environments than around isolated galaxies, especially at Cosmic Noon.

\section{Conclusions}
\label{sec:conclusions}

As part of our larger programme to identify the host galaxies of known {\MgII} and {\CIV} absorbers at Cosmic Noon ($z=2-3$) with KCWI \citep[e.g.][]{nielsen20}, we identified a compact group of four galaxies giving rise to a DLA with ultra-strong {\MgII} absorption at $z=2.43$ in quasar field J234628$+$124859. A previous survey of the field within $D\sim12.5$~kpc of the quasar sightline yielded no host galaxies with {\Halpha} emission to a SFR limit of $<4-11$~M$_{\odot}$~yr$^{-1}$ \citep{wang2015}, in contrast to our wider and more sensitive survey out to $D<140$~kpc. We compared the properties of these galaxies to the kinematically complex, multiphase absorption using detailed photoionisation modelling. Our results include the following.

\begin{enumerate}
    \item[(1) {\it Compact Group Properties}:] The compact group galaxies are located within $D=20-40$~kpc of the background quasar sightline, with a group centre impact parameter of $D\sim25$~kpc, which is located beyond the search radius of the previous survey. Three of the galaxies (G1, G2, and G3) were identified via their {\Lya} emission while the fourth galaxy (G4) did not have {\Lya} emission so it was identified by its ISM absorption lines (Fig.~\ref{fig:galaxies}). All four galaxies are located within $D=8-28$~kpc and $v\sim40-340$~{\kms} of each other, where interactions between the galaxies are likely considerable. In fact, G1 and G4 appear to have narrow tidal features in the {\it HST} imaging (Fig.~\ref{fig:whitelight}). G1 dominates the group, where it is an $1.66L_{\ast}$ galaxy, and the other three galaxies are less massive, with $\sim0.1-0.3L_{\ast}$ (Tables~\ref{tab:HSTgalprops}, \ref{tab:KCWIgalprops}). While the galaxies are likely interacting, none of the galaxies is undergoing a starburst. G1 and G2 may be capable of driving outflows with ${\rm SFR}_{\rm FUV}=11.6$ and $1.7$~M$_{\odot}$~yr$^{-1}$, respectively, corresponding to $\Sigma_{\rm SFR}\sim0.07$~M$_{\odot}$~yr$^{-1}$~kpc$^{-2}$ (these values may be larger if we assume more dust extinction). In contrast, G3 and G4 are unlikely to drive significant outflows with SFRs of $1.4$ and $2.0$~M$_{\odot}$~yr$^{-1}$, respectively, corresponding to $\Sigma_{\rm SFR}\sim0.006$~M$_{\odot}$~yr$^{-1}$~kpc$^{-2}$. 
    
    \item[(2) {\it Absorption Properties}:] The absorption along the quasar sightline is kinematically complex, with the bulk of the metal-line absorption spanning a velocity width of $\Delta v\sim 450$~{\kms} and additional highly blueshifted weak absorption at $\sim-400$~{\kms} for a total width of $\Delta v=700$~{\kms} (Fig.~\ref{fig:absorption}). This large kinematic spread is consistent with other ultra-strong {\MgII} absorbers that were interpreted as superwinds from starbursts. The redshifts of the four galaxies align with the bulk of the metal-line absorption, but the highly blueshifted clouds are offset by at least $250$~{\kms} from the host galaxies.
    
    \item[] We used the cloud-by-cloud multiphase Bayesian modelling method \citep[CMBM;][]{sameer2021, sameerleo} to probe the various structures and ionisation phases of the gas along the quasar sightline. The 30 modelled clouds (Table~\ref{tab:cmbm}) across the low ionisation phase constrained on {\HI}, {\MgII}, {\FeII}, or {\SiII} (20 clouds) and the intermediate ionisation phase constrained on {\SiIV} or {\CIV} (10 clouds) yielded a wide range of properties (see Fig.~\ref{fig:violins}). The cloud column densities are as low as typical values inferred for {\Lya} forest absorbers and as large as nearly DLA strengths ($13\lesssim\log(N({\HI})/{\rm cm}^{-2})\lesssim20$), where the total column density is $\log(N({\HI})/{\rm cm}^{-2})=20.53$ (20.34 for the low ionisation phase and 20.07 for the intermediate ionisation phase). We also inferred a wide range of cloud metallicities ($-3\lesssim\log(Z/Z_{\odot})\lesssim1$, where both phases roughly have the same range), corresponding to IGM metallicities on the low end and super-solar metallicities, which are above average ISM values at this epoch, on the high end. These wide ranges of cloud properties suggest that the average values typically reported for such kinematically-complex systems neglect the diversity of physical structures probed along the line-of-sight.
        
    \item[] Our metallicity modelling inferred an $N({\HI})$-weighted mean metallicity of $\log(Z/Z_{\odot})=-0.68$ (the low ionisation phase has $-0.57$ and the intermediate ionisation phase has $-0.93$; Fig.~\ref{fig:ZNHI}). This value is consistent with previous estimates for this system, suggesting that the CMBM method recovers the average physical properties of the system while also making the realistic assumption that the ionisation conditions vary across the different structures being probed. Compared to the wider DLA literature, this system has a lower metallicity than expected for its full metal line width, which may be a result of the group environment and/or significant amounts of metal-poor accretion.
    
    \item[(3) {\it Possible Gas Origins}:] The highly blueshifted, metal-rich clouds likely trace outflowing gas from G1, which may have an inclination-corrected outflow velocity as large as $v_{\rm out}=1000$~{\kms} ($i=66^{\circ}$ for G1). This velocity is comparable to the values inferred from stacked galaxy spectra at this epoch \citep[e.g.][]{steidel10} and is the upper bound set for the outflow in the first system we studied for our $z\sim2$ KCWI survey \citep{nielsen20}. Since G2 may also be capable of driving outflows with a lower SFR, other metal-rich clouds at lower velocity (corresponding to inclination-corrected velocities of $v_{\rm out}=230-570$~{\kms}) may also be outflowing gas.
    
    \item[] We attributed very metal-poor ($\log(Z/Z_{\odot})\lesssim-2$) clouds at low velocity relative to galaxies G1, G2, and G4 as being accretion from the IGM. Their metallicities are comparable to typical {\Lya} forest systems \citep[e.g.][]{simcoe04} and they have very large column densities that are nearly DLAs, consistent with predictions from simulations \citep[e.g.][]{vandevoort+schaye12}. Given the group environment and the low velocity of these clouds relative to the galaxies (i.e. lower than typical rotation velocities), it is likely that this gas is accreting onto the group environment, where it is currently unclear if this material reaches the individual group member galaxies as cold mode accretion.
    
    \item[] The rest of the clouds with metallicities typical of the ISM and CGM at this epoch likely trace tidal streams from the strong galaxy--galaxy interactions expected in compact group environments. These clouds have velocities comparable to all four host galaxy redshifts, suggesting that they may trace the wider gravitational potential well of the group and will eventually become part of the wider IGrM to later recycle back onto the galaxies.
    
\end{enumerate}

In contrast to low redshift and nearby surveys of compact groups, we find evidence of not only the tidal streams between galaxies commonly observed in such systems, but also outflows with large velocities and a considerable amount of accretion onto the wider group environment. This compact group may be in an earlier evolutionary phase \citep[e.g.][]{verdes-montenegro01}, where at least two of the galaxies may still be accreting the gas necessary to sustain their star formation. To better place this system in context with lower redshift work, it would be ideal to determine how {\HI}-deficient the galaxies are relative to their SFRs and to map the CGM in emission to determine if the quasar directly probes a tidal stream, and if so, from which galaxy the stream originates. ISM metallicities of the galaxies would also be useful for better confirming tidal material, but we do not have coverage of the emission lines required for this measurement in the KCWI data. On the other end, low redshift observations of quasar sightlines through compact groups in the various evolutionary phases are needed to obtain a more complete census of the IGrM reservoir since these sightlines can probe more diffuse and more highly ionised gas than {\HI} studies. These data would also confirm if the highly complex absorption is a feature of compact groups or simply the nature of gas flows at Cosmic Noon. Regardless, compact groups and small group environments in general may be more likely to host rare systems like DLAs and ultra-strong {\MgII} systems given the complexity of galaxy--galaxy interactions, especially at low impact parameters.

\section*{Acknowledgements}

N.M.N.~thanks D.~B.~Fisher, S.~Cantalupo, and D.~Rupke for discussions related to the illumination gradient present in the pipeline-reduced KCWI data and PSF subtraction. We also thank the referee for their helpful comments which have improved the paper.

N.M.N., G.G.K., and M.T.M.~acknowledge the support of the Australian Research Council through {\it Discovery Project} grant DP170103470. Parts of this research were supported by the Australian Research Council Centre of Excellence for All Sky Astrophysics in 3 Dimensions (ASTRO 3D), through project number CE170100013. C.W.C. and J.C.C. were supported by the National Science Foundation through grant NSF AST-1517816 and by NASA through HST grant GO-13398 from the Space Telescope Science Institute, which is operated by the Association of Universities for Research in Astronomy, Inc., under NASA contract NAS5-26555.

This research made use of Montage. It is funded by the National Science Foundation under Grant Number ACI-1440620, and was previously funded by the National Aeronautics and Space Administration's Earth Science Technology Office, Computation Technologies Project, under Cooperative Agreement Number NCC5-626 between NASA and the California Institute of Technology.

Some of the data presented herein were obtained at the W. M. Keck Observatory, which is operated as a scientific partnership among the California Institute of Technology, the University of California and the National Aeronautics and Space Administration. The Observatory was made possible by the generous financial support of the W. M. Keck Foundation. Observations were supported by Swinburne Keck programs with KCWI: 2017B\_W270, 2018A\_W185, and 2018B\_W232; with NIRC2-LGS: 2019B\_W237; and NASA Keck program with KCWI: 2020B\_N021. The authors wish to recognise and acknowledge the very significant cultural role and reverence that the summit of Maunakea has always had within the indigenous Hawaiian community. We are most fortunate to have the opportunity to conduct observations from this mountain.

\section*{Data Availability}

The reduced VLT/UVES quasar spectra are available in the SQUAD database \citep{uvessquad} while the {\it HST} imaging is available via the Hubble Legacy Archive.\footnote{\url{https://hla.stsci.edu/}} The reduced KCWI data underlying this article will be shared on reasonable request to the corresponding author.



\bibliographystyle{mnras}
\bibliography{refs} 



\appendix

\section{KCWI Illumination Gradient and Scattered Light Correction}
\label{app:gradient}

\begin{figure}
\includegraphics[width=\linewidth]{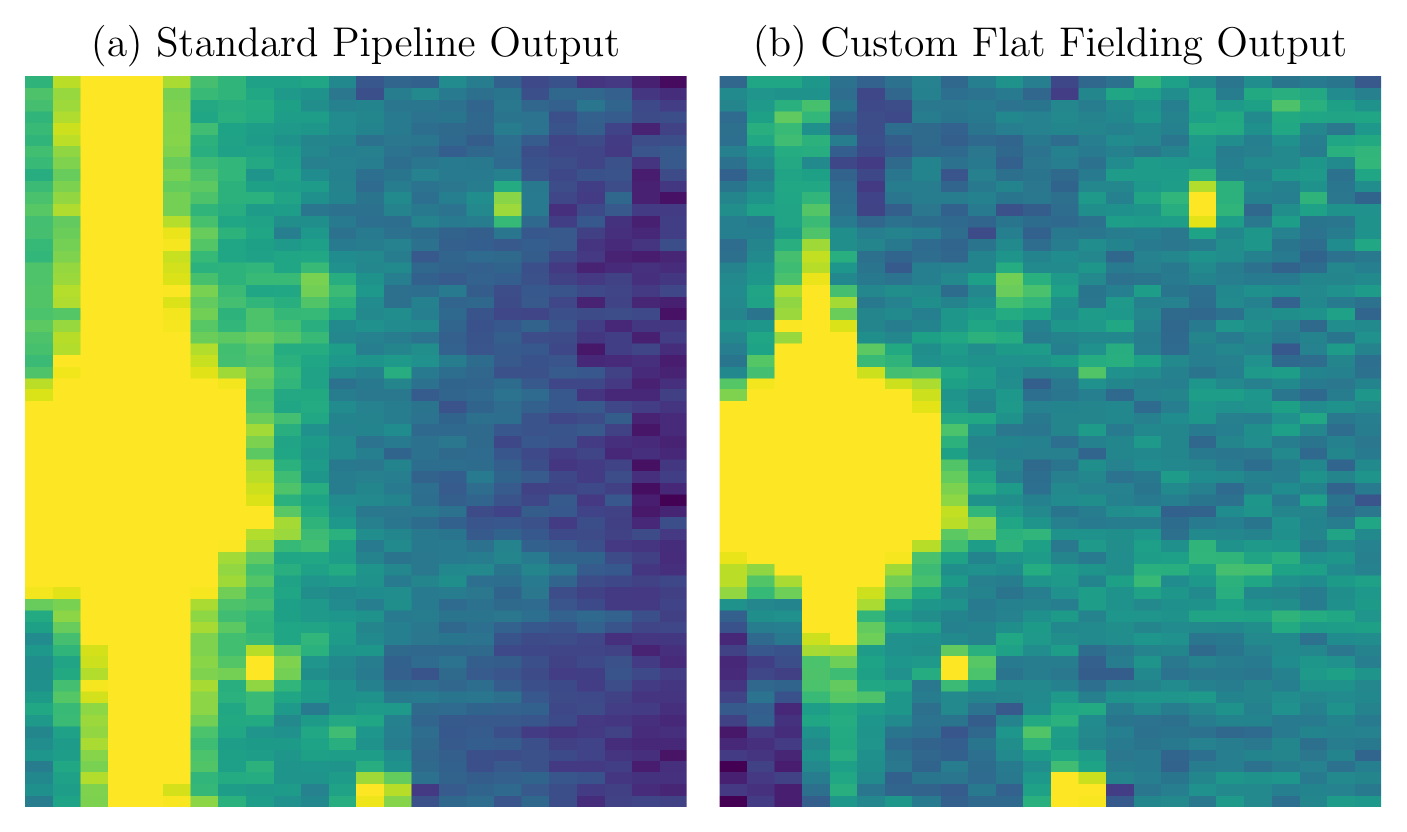}
\caption{White light images for a single exposure using (a) the standard KCWI DRP method and (b) our custom illumination gradient and scattered light correction. Both the gradient from left to right and the scattered light vertically in the same slices as the quasar are removed.}
\label{fig:flatwhite}
\end{figure}

The KCWI DRP (IDL, v1.2.2) does not fully remove a ten percent slice-by-slice wavelength-dependent illumination gradient in the data.\footnote{As of publication, we have not yet tested the python version of the pipeline.} The pipeline also does not adequately correct for scattered light in the slices containing the bright background quasar, which makes these slices brighter than the rest. Fig.~\ref{fig:flatwhite} demonstrates these effects generally (i.e. the wavelength-dependence is not reflected here), where panel (a) is a white light image of a single exposure for the J2346$+$124 field for the cube output from the pipeline with default settings. A strong gradient is present from left to right across the 24 slices and this gradient is present regardless of the field being observed or whether a bright object such as a quasar is within the FOV. The gradient is always left-to-right in the data regardless of the orientation of the FOV on the sky, so it is not a sky position angle effect. The bright slices on the left also have significant scattered light due to the quasar centred vertically in the field; if the quasar is located on the right side of the FOV, then those slices would instead have the bright scattered light. We have found these effects in all combinations of the large and medium slicers with the BL and BM gratings, as well as the small slicer with the BL grating. They have also been documented by other KCWI users \citep[e.g.][]{cai19, rupkemakani}.

\begin{figure}
	\includegraphics[width=\linewidth]{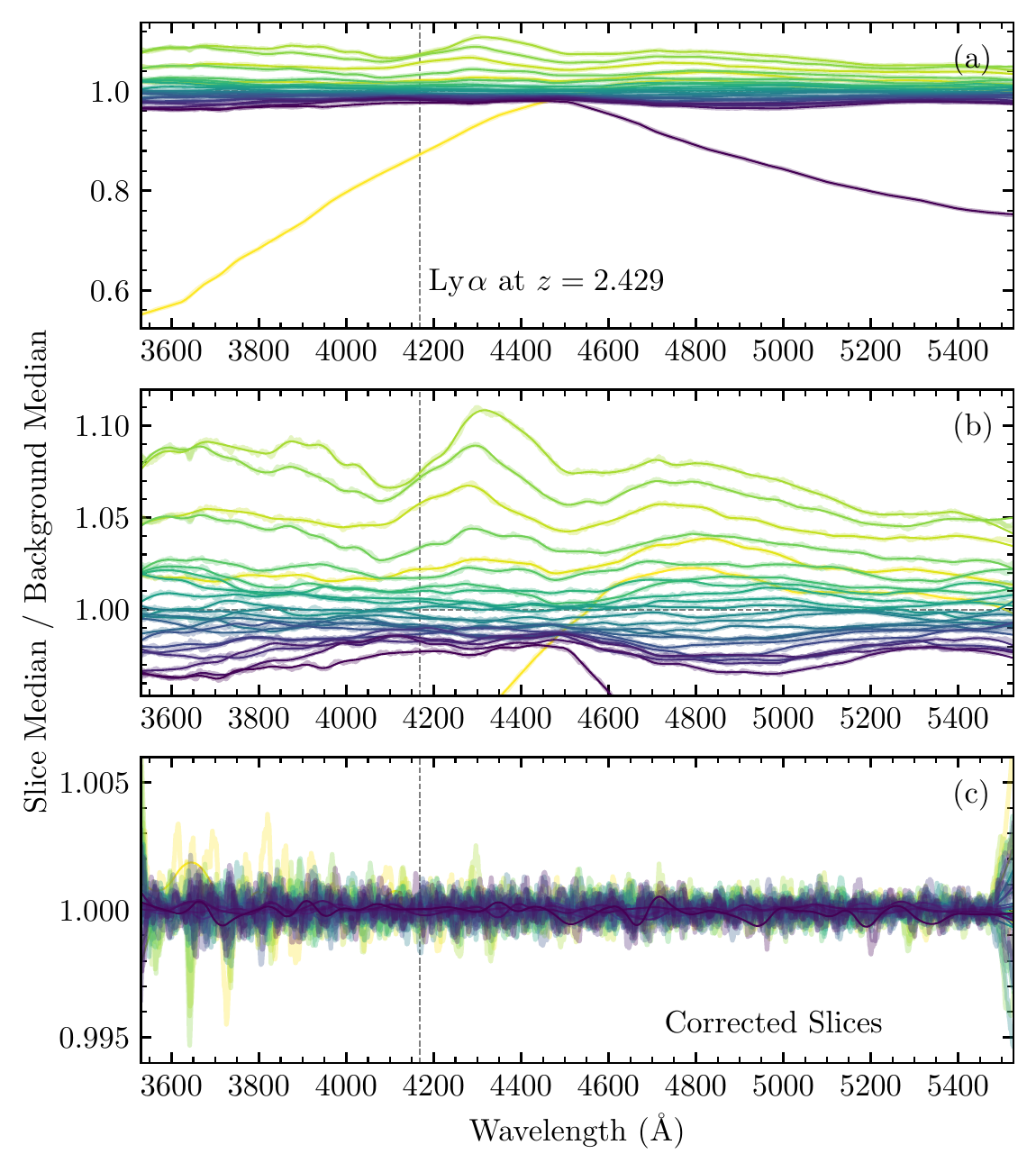}
    \caption{Slice-by-slice wavelength-dependent illumination gradient and scattered light in the KCWI data. Each curve represents each of the 24 slices, where similar-coloured slices are neighbours on the CCD. The median spectrum in each slice is compared to the median spectrum of the background. The quasar and continuum objects have been masked out for this comparison. Thin lines are a spline to the boxcar-smoothed data (thick fainter lines) which had been re-binned into 40~{\AA} wide bins. The vertical line at $\lambda\sim4170$~{\AA} represents the wavelength at which {\Lya} emission is located for the DLA. (a) Full range of the gradient. Edge slices are represented by the yellow and purple curves that decrease to very low ratios at low and high wavelength, respectively. (b) A zoom-in on the majority of the slices to show the range of the deviation from a flat illumination pattern. Slices that contain the quasar are represented by the highest ratio green and yellow curves and the peak of these curves at $\lambda\sim4350$~{\AA} corresponds to {\Lya} at $z_{\rm qso}$, indicating that scattered light causes a significant wavelength-dependent deviation from the rest of the slices in the cubes. (c) Slices after correcting for the illumination gradient and scattered light, demonstrating that the method results in cubes that are flat to less than 1 percent.}
    \label{fig:gradient}
\end{figure}

To remedy these issues, we determine the gradient across the field and then divide it out of the cube. Ideally separate sky fields would be used for this to avoid removing or correcting out the desired {\Lya} signal or other continuum signal, but we do not obtain separate sky fields for this program since there is an adequate amount of sky in each pointing. To measure the gradient, we use non-sky-subtracted, flux-calibrated cubes from the reduction pipeline by skipping step {\texttt{kcwi\_stage5sky}}. This method makes the assumption that the sky dominates the signal in the cube and that the sky is constant across the FOV, so any deviations from a flat background are likely due to instrumental illumination effects that are not removed by the pipeline. 

Measuring the gradient requires obtaining only the sky background and not including any objects in the estimate. Given this, we developed an automated object masking routine that uses sigma-clipping to mask both the significant quasar light and other objects in the field. This method can run without first identifying objects in the field by the user. For this method, we created a white light image spanning $3500\lesssim\lambda\lesssim5500$~{\AA} for each flux-calibrated, non-sky-subtracted exposure output from the pipeline. This image was then used to automatically identify objects in the field. When present, the quasar was masked as the first step using a stringent $0.02\sigma$ sigma-clipping to remove the majority of the non-sky signal. From the quasar-masked image we then further masked any continuum objects slice-by-slice using a $2\sigma$ sigma-clipping, running this step twice to ensure adequate masking. This continuum object masking step needs to be done slice-by-slice due to the gradient and quasar scattered light, which impact the median and standard deviation of the flux across the cube. The sigma-clipping thresholds are not necessarily fixed for all fields in our survey; rather, they depend on both the quasar brightness (since each quasar in our survey has a different spectrum/luminosity) and the background (since a brighter background will require a lower threshold to remove faint galaxies). However, the thresholds are held constant for all exposures in a given field. The remaining unmasked spaxels, which define the background or sky, were then used to measure and correct the illumination gradient.

\begin{figure*}
    \centering
    \includegraphics[width=\linewidth]{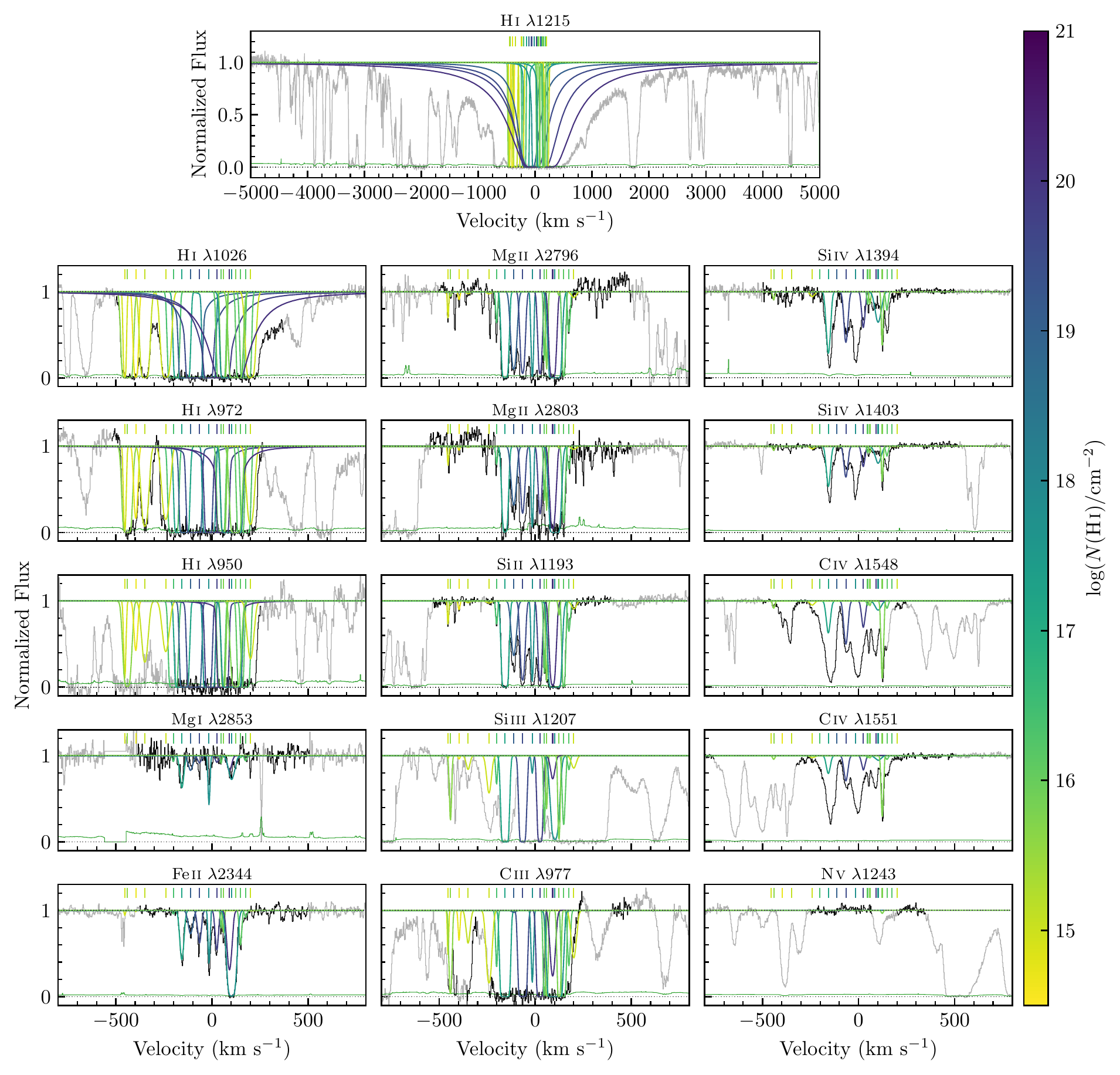}
    \caption{Components constrained with the low ionisation lines {\HI}, {\MgII}, {\FeII}, or {\SiII}, where the ticks above the spectrum indicate the component central velocity ($v=0$~{\kms} represents $z_{\rm abs}$ measured from {\MgII}). VP component colours represent the {\HI} column density. The low ionisation lines are well-fitted by these components, which account for the vast majority of their absorption, but the intermediate ionisation lines are significantly under-fit in this single phase model. Additional components are required for {\SiIII}, {\SiIV}, and {\CIV} as shown in Fig.~\ref{fig:highions}. The total fit representing both these components and the intermediate ionisation components (Fig.~\ref{fig:highions}) is shown in Fig.~\ref{fig:absorption}.}
    \label{fig:lowions}
\end{figure*}

As a reference to compare each slice to, we calculated the median background spectrum for the entire FOV using the unmasked spaxels. This background spectrum was smoothed by convolving with a boxcar filter with a kernel size of 201 spectral pixels, and then this smoothed spectrum was re-binned into 40~{\AA} wide bins. This binned spectrum was then interpolated onto the original wavelength array using a spline. These steps smoothed out any bright emission lines from the sky or galaxies in the field as well as objects with faint continuum emission not caught in the masking step, all of which would bias the gradient measurement. They also reduced the effect of noise since we are only interested in broader illumination differences between slices. Then the median, smoothed spectrum for each slice was measured using the same steps and compared to the median, smoothed background reference spectrum. This comparison, the ratio between the slice median and the background median, is plotted in Fig.~\ref{fig:gradient} for the J2346$+$124 exposure shown in Fig.~\ref{fig:flatwhite}.

\begin{figure*}
    \centering
    \includegraphics[width=\linewidth]{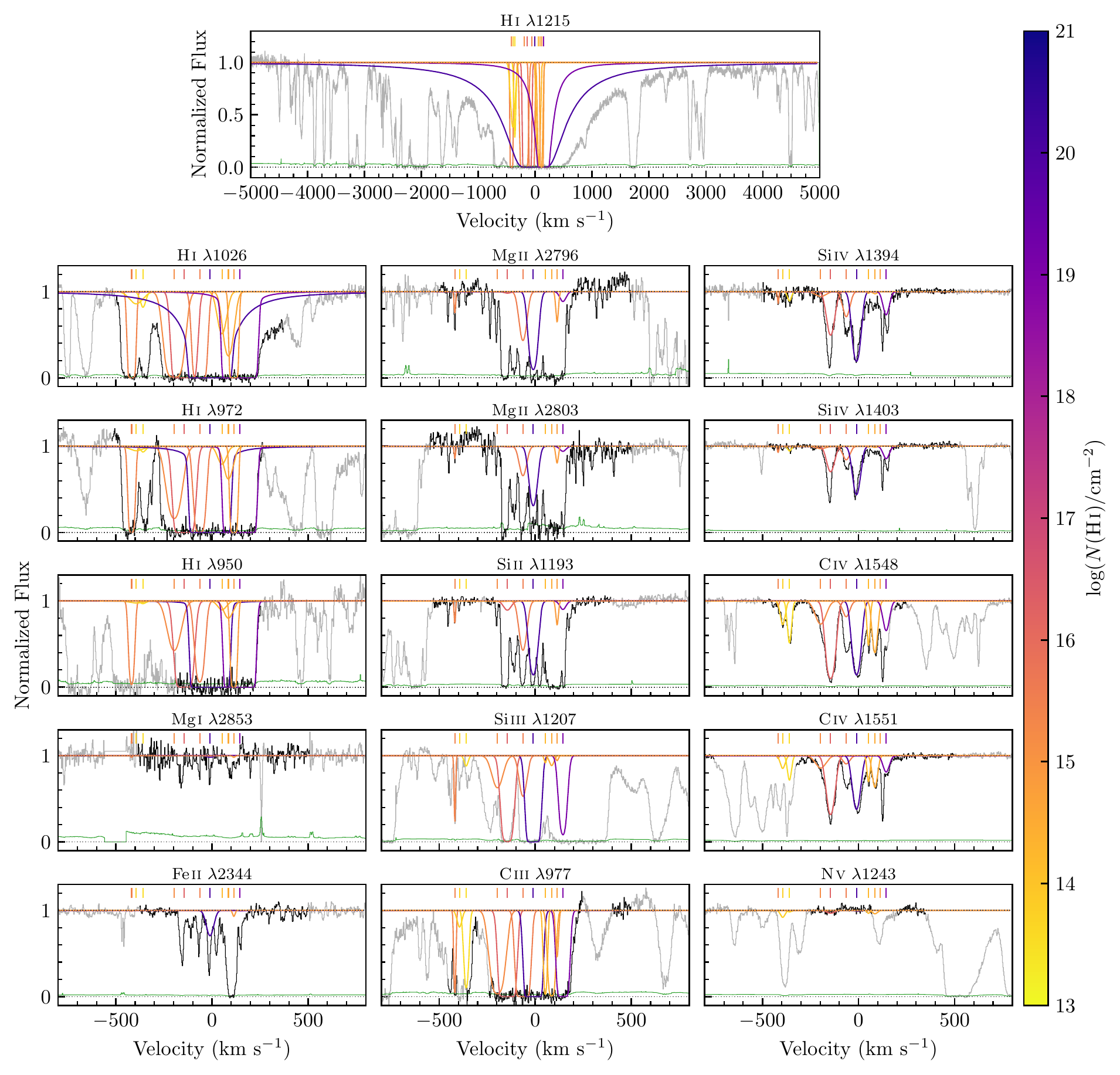}
    \caption{Components constrained with the intermediate ionisation lines {\SiIV} or {\CIV}, where the ticks above the spectrum indicate the component central velocity ($v=0$~{\kms} represents $z_{\rm abs}$ measured from {\MgII}). VP component colours represent the {\HI} column density. The intermediate ionisation lines are well-fitted by these components, which account for the vast majority of their absorption, but do not fit the low ionisation lines well. The total fit representing these components and the low ionisation components (Fig.~\ref{fig:lowions}) is shown in Fig.~\ref{fig:absorption}.}
    \label{fig:highions}
\end{figure*}

Figs.~\ref{fig:gradient}(a) and (b) quantify the wavelength-dependent illumination gradient effect by showing the ratio of the median spectrum for each of the 24 slices (objects are masked out) to the median background spectrum. Brighter slices in the gradient are represented by yellow curves while fainter slices are purple, where neighbouring slices have similar colours. Slices with the largest slice-to-background ratios represent those slices in which the quasar is contained and are the result of scattered light from the bright quasar. For these slices, the peak in the slice-to-background ratio at $\lambda\sim4350$~{\AA} corresponds to {\Lya} from the quasar at $z_{\rm qso}$. We also highlight the wavelength where we expect to find {\Lya} emission from DLA-host galaxies to demonstrate that we are not removing any signal from this system (i.e. there is no peak in the ratios that will be over-subtracted from the cube). The two slices that have very low ratios at lower wavelengths (yellow) and higher wavelengths (purple) are on the edges of the CCD. This plot shows that the gradient across all 24 slices is on the order of more than ten percent for the lowest wavelengths and less for the highest wavelengths, with an additional increase in the gradient at wavelengths $4600\lesssim\lambda\lesssim5100$~{\AA}. 

To correct the cubes for the illumination gradient, these wavelength-dependent ratios were divided out spaxel-by-spaxel for each slice. The corrected slice ratios are plotted in Fig.~\ref{fig:gradient}(c), showing that the background becomes flat to within one percent using this method. To finalise the reduction, we measured a median sky in the corrected cube using only the non-masked spaxels and subtracted this sky from every spaxel since the sky subtraction step of the pipeline was skipped. This resulted in adequate flat-fielding, scattered light correction, and sky subtraction in the final data cubes. The result of this correction is shown in Fig.~\ref{fig:flatwhite}(b), where faint features and even galaxies on the right side of the FOV are more prominent. The fully reduced and flat-fielded cubes were then combined according to the steps laid out in Section~\ref{sec:KCWIdata}.

\begin{table*}
    \centering
    \caption{Cloud-by-Cloud Multiphase Bayesian Modelling Results}
    \label{tab:cmbm}
    \begin{tabular}{lrrrrrrr}
    \hline
    Constraining & $v$       & $\log N({\HI})$ & $b_{\rm turb}$ & $\log n_{\rm H}$ & $\log Z$      & $\log T$ & $\log L$ \\
    Ion        & ({\kms})  & (cm$^{-2}$)     & ({\kms})       & (cm$^{-3}$)      & ($Z_{\odot}$) & (K)      & (pc)     \\
    \hline
    \multicolumn{8}{c}{Components Constrained on Low Ionisation Lines} \\
    \hline
    {\SiII}\_0   & $-454$ & $14.94_{-0.07}^{+0.11}$ &  $ 0.1_{- 0.1}^{+ 0.6}$ &  $-1.40_{-0.11}^{+0.19}$ &  $ 0.64_{-0.08}^{+0.12}$ & $3.25_{-0.02}^{+0.02}$ & $-0.85_{-0.13}^{+0.12}$  \\[3pt]
    {\SiII}\_1   & $-441$ & $15.66_{-0.08}^{+0.05}$ &  $ 4.6_{- 1.0}^{+ 0.1}$ &  $-1.97_{-0.10}^{+0.05}$ &  $-1.04_{-0.11}^{+0.06}$ & $4.26_{-0.01}^{+0.01}$ & $ 1.66_{-0.04}^{+0.04}$  \\[3pt]
    {\SiII}\_3   & $-397$ & $14.71_{-0.08}^{+0.04}$ &  $ 3.6_{- 1.8}^{+ 1.3}$ &  $-1.27_{-0.11}^{+0.14}$ &  $ 0.09_{-0.07}^{+0.12}$ & $3.82_{-0.04}^{+0.04}$ & $-0.98_{-0.15}^{+0.14}$  \\[3pt]
    {\HI}\_0     & $-350$ & $15.09_{-0.03}^{+0.03}$ &  $12.2_{- 2.7}^{+ 3.4}$ &  $-1.56_{-0.35}^{+1.85}$ &  $-0.57_{-1.36}^{+0.29}$ & $4.12_{-0.07}^{+0.07}$ & $ 0.18_{-0.76}^{+0.68}$  \\[3pt]
    {\HI}\_3     & $-241$ & $14.99_{-0.66}^{+0.57}$ &  $14.8_{-14.8}^{+18.1}$ &  $-2.02_{-0.86}^{+0.46}$ &  $-0.34_{-1.76}^{+0.89}$ & $4.16_{-0.35}^{+0.43}$ & $ 0.98_{-1.10}^{+1.63}$  \\[3pt]
    {\MgII}\_0   & $-202$ & $16.36_{-0.55}^{+1.07}$ &  $ 4.0_{- 2.6}^{+ 2.2}$ &  $-0.87_{-0.61}^{+1.39}$ &  $-0.76_{-1.00}^{+0.96}$ & $4.05_{-0.09}^{+0.08}$ & $ 0.09_{-1.06}^{+1.05}$  \\[3pt]
    {\FeII}\_1   & $-158$ & $17.19_{-0.11}^{+0.26}$ &  $12.6_{- 0.3}^{+ 0.4}$ &  $-1.54_{-0.03}^{+0.03}$ &  $-0.07_{-0.17}^{+0.07}$ & $3.93_{-0.02}^{+0.03}$ & $ 2.10_{-0.08}^{+0.09}$  \\[3pt]
    {\FeII}\_3   & $-112$ & $19.12_{-1.68}^{+0.77}$ &  $15.6_{- 2.8}^{+ 3.3}$ &  $ 0.86_{-1.10}^{+0.14}$ &  $-1.49_{-0.76}^{+1.48}$ & $3.83_{-0.17}^{+0.14}$ & $-0.24_{-1.42}^{+1.45}$  \\[3pt]
    {\SiII}\_7   & $ -66$ & $19.51_{-0.02}^{+0.01}$ &  $11.9_{- 0.2}^{+ 0.4}$ &  $-2.10_{-0.01}^{+0.01}$ &  $-1.96_{-0.01}^{+0.02}$ & $4.25_{-0.01}^{+0.01}$ & $ 4.11_{-0.01}^{+0.01}$  \\[3pt]
    {\SiII}\_9   & $ -16$ & $17.76_{-0.08}^{+0.11}$ &  $ 4.5_{- 0.1}^{+ 0.1}$ &  $-0.14_{-0.09}^{+0.15}$ &  $-0.49_{-0.07}^{+0.06}$ & $3.93_{-0.02}^{+0.02}$ & $-0.05_{-0.20}^{+0.15}$  \\[3pt]
    {\SiII}\_10  & $ +24$ & $19.76_{-0.02}^{+0.03}$ &  $10.1_{- 0.2}^{+ 0.2}$ &  $-2.02_{-0.01}^{+0.01}$ &  $-2.11_{-0.01}^{+0.01}$ & $4.20_{-0.01}^{+0.01}$ & $ 3.99_{-0.03}^{+0.03}$  \\[3pt]
    {\SiII}\_12  & $ +47$ & $16.09_{-0.44}^{+1.13}$ &  $ 2.2_{- 1.1}^{+ 0.6}$ &  $-1.44_{-0.18}^{+0.55}$ &  $-0.02_{-0.88}^{+0.34}$ & $3.91_{-0.09}^{+0.10}$ & $ 0.78_{-0.15}^{+0.16}$  \\[3pt]
    {\SiII}\_13  & $ +59$ & $15.76_{-0.07}^{+0.25}$ &  $ 4.1_{- 0.7}^{+ 0.1}$ &  $-1.62_{-0.07}^{+0.13}$ &  $ 0.47_{-0.29}^{+0.14}$ & $3.50_{-0.23}^{+0.26}$ & $ 0.52_{-0.14}^{+0.14}$  \\[3pt]
    {\FeII}\_7   & $ +90$ & $20.07_{-0.34}^{+0.23}$ &  $14.6_{- 2.9}^{+ 0.1}$ &  $-0.95_{-0.61}^{+1.76}$ &  $-1.92_{-0.31}^{+0.32}$ & $3.98_{-0.02}^{+0.02}$ & $ 2.63_{-0.27}^{+0.27}$  \\[3pt]
    {\FeII}\_8   & $+102$ & $17.63_{-0.21}^{+0.16}$ &  $20.3_{- 0.6}^{+ 0.7}$ &  $-1.34_{-0.07}^{+0.04}$ &  $ 0.25_{-0.10}^{+0.25}$ & $3.69_{-0.06}^{+0.06}$ & $ 1.86_{-0.05}^{+0.05}$  \\[3pt]
    {\FeII}\_9   & $+113$ & $15.81_{-4.81}^{+3.70}$ &  $ 7.6_{- 7.7}^{+ 1.2}$ &  $-0.90_{-3.10}^{+1.90}$ &  $-0.06_{-2.94}^{+1.03}$ & $3.78_{-0.63}^{+0.78}$ & $-0.71_{-6.62}^{+4.32}$  \\[3pt]
    {\FeII}\_10  & $+123$ & $15.82_{-1.82}^{+0.75}$ &  $ 0.3_{- 0.3}^{+ 4.2}$ &  $-2.49_{-0.39}^{+0.15}$ &  $-0.87_{-0.68}^{+1.47}$ & $4.41_{-0.04}^{+0.03}$ & $ 2.99_{-0.39}^{+0.38}$  \\[3pt]
    {\SiII}\_17  & $+148$ & $16.14_{-0.04}^{+0.12}$ &  $ 7.2_{- 0.3}^{+ 0.1}$ &  $-1.58_{-0.04}^{+0.07}$ &  $ 0.41_{-0.14}^{+0.10}$ & $3.59_{-0.06}^{+0.06}$ & $ 0.90_{-0.04}^{+0.04}$  \\[3pt]
    {\SiII}\_18  & $+175$ & $16.21_{-0.64}^{+1.54}$ &  $ 7.1_{- 1.4}^{+ 1.4}$ &  $-0.80_{-0.28}^{+0.39}$ &  $-0.31_{-1.34}^{+0.44}$ & $3.94_{-0.05}^{+0.06}$ & $-0.30_{-0.14}^{+0.15}$  \\[3pt]
    {\SiII}\_19  & $+200$ & $15.06_{-0.65}^{+0.65}$ &  $16.1_{- 5.8}^{+ 5.1}$ &  $-1.52_{-0.47}^{+0.41}$ &  $ 0.02_{-0.49}^{+0.58}$ & $3.91_{-0.13}^{+0.11}$ & $-0.11_{-0.40}^{+0.35}$  \\[3pt]
    \hline
    \multicolumn{8}{c}{Components Constrained on Intermediate Ionisation Lines} \\
    \hline
    {\SiIV}\_1   & $-419$ & $15.41_{-0.08}^{+0.12}$ &  $ 0.3_{- 0.3}^{+1.5}$ &  $-1.73_{-0.05}^{+0.06}$ &  $-0.25_{-0.11}^{+0.07}$ & $4.06_{-0.01}^{+0.01}$ & $0.77_{-0.06}^{+0.07}$  \\[3pt]
    {\SiIV}\_2   & $-395$ & $13.78_{-0.10}^{+0.15}$ &  $ 9.6_{- 2.3}^{+0.2}$ &  $-3.40_{-0.03}^{+0.09}$ &  $-1.01_{-0.13}^{+0.10}$ & $5.11_{-0.02}^{+0.01}$ & $3.86_{-0.07}^{+0.06}$  \\[3pt]
    {\SiIV}\_3   & $-360$ & $13.52_{-0.21}^{+0.48}$ &  $12.6_{- 0.8}^{+1.0}$ &  $-2.94_{-0.06}^{+0.10}$ &  $ 0.54_{-0.44}^{+0.20}$ & $4.14_{-0.08}^{+0.09}$ & $1.34_{-0.15}^{+0.19}$  \\[3pt]
    {\CIV}\_0    & $-197$ & $15.27_{-1.71}^{+1.08}$ &  $36.4_{-10.6}^{+8.5}$ &  $-2.67_{-0.36}^{+0.24}$ &  $-0.85_{-1.26}^{+1.36}$ & $4.49_{-0.13}^{+0.13}$ & $2.88_{-1.01}^{+0.93}$  \\[3pt]
    {\CIV}\_2    & $-146$ & $16.32_{-0.47}^{+0.32}$ &  $26.3_{- 1.3}^{+2.0}$ &  $-2.57_{-0.06}^{+0.06}$ &  $-0.97_{-0.41}^{+0.50}$ & $4.46_{-0.05}^{+0.05}$ & $3.71_{-0.34}^{+0.34}$  \\[3pt]
    {\SiIV}\_4   & $ -65$ & $15.56_{-0.03}^{+0.02}$ &  $24.6_{- 1.6}^{+0.4}$ &  $-2.22_{-0.02}^{+0.02}$ &  $ 0.98_{-0.02}^{+0.02}$ & $3.20_{-0.01}^{+0.01}$ & $1.33_{-0.04}^{+0.03}$  \\[3pt]
    {\SiIV}\_5   & $ -11$ & $20.03_{-0.01}^{+0.01}$ &  $24.9_{- 0.1}^{+0.1}$ &  $-2.48_{-0.01}^{+0.01}$ &  $-2.21_{-0.01}^{+0.01}$ & $4.41_{-0.01}^{+0.01}$ & $5.20_{-0.01}^{+0.01}$  \\[3pt]
    {\CIV}\_8    & $ +52$ & $14.23_{-0.87}^{+1.52}$ &  $ 3.4_{- 3.5}^{+3.0}$ &  $-3.10_{-0.33}^{+0.43}$ &  $-0.79_{-1.54}^{+0.91}$ & $4.73_{-0.21}^{+0.21}$ & $3.03_{-0.69}^{+0.68}$  \\[3pt]
    {\CIV}\_9    & $ +85$ & $14.56_{-1.40}^{+0.47}$ &  $15.1_{- 2.7}^{+1.9}$ &  $-3.08_{-0.28}^{+0.19}$ &  $-0.69_{-0.53}^{+1.47}$ & $4.69_{-0.17}^{+0.15}$ & $3.24_{-0.59}^{+0.51}$  \\[3pt]
    {\CIV}\_11   & $+143$ & $18.99_{-0.18}^{+0.50}$ &  $21.2_{- 2.0}^{+1.3}$ &  $-2.45_{-0.01}^{+0.02}$ &  $-2.99_{-0.01}^{+0.11}$ & $4.41_{-0.01}^{+0.01}$ & $5.14_{-0.04}^{+0.03}$  \\[3pt]
    \hline
    \end{tabular}
\end{table*}

\section{Inferred Cloud Properties from Photoionisation Modelling with CMBM}
\label{app:metallicity}

As detailed in Section~\ref{sec:absorption}, we required 30 different VP components across two ionisation phases to adequately model the absorption profile for this DLA system. The full list of ions used to constrain the models, where the doublets and multiplets were covered and modelled, have their column densities tabulated in Table~\ref{tab:columns}. Fig.~\ref{fig:lowions} plots several representative transitions superimposed with the individual VP components for the 20 components that were constrained using a low ionisation line, including {\HI}, {\MgII}, {\FeII}, or {\SiII}. Most components were constrained on {\SiII}, while two components were constrained on {\HI}, where no metal lines required a component. An additional 10 components were required in the {\SiIV} and {\CIV} lines to define the intermediate ionisation phase and are plotted in Fig.~\ref{fig:highions}. In both cases, the components are coloured by their {\HI} column density, where darker colours represent higher column densities. Each component is applied to the full suite of transitions used in the metallicity modelling and this applies constraints on the {\sc Cloudy} grids.

Table~\ref{tab:cmbm} lists the properties for each of the 30 components modelled to the system, including the (1) ion with which each component was constrained on, where the names indicate the ion (e.g. {\SiII} or {\CIV}) and an identification number (e.g. ion\_0) for reference, (2) central velocity, (3) {\HI} column density, (4) Doppler $b$-parameter, (5) hydrogen number density, (6) metallicity, (7) temperature, and (8) thickness. The values reported are the median of the posterior distribution, while the errors represent the minimum and maximum of the distributions. This is in contrast to most literature, which often report $1\sigma$ uncertainties, so the uncertainties reported here appear to be larger than usual for a subset of components. The components are separated by whether they were constrained on the low ionisation lines or the intermediate ionisation lines for convenience. The full posterior distributions, which have been sigma-clipped at the $3\sigma$ level for clarity, are presented in Fig.~\ref{fig:violins}.


\bsp	
\label{lastpage}
\end{document}